\documentstyle[aps,twocolumn,epsfig,eqsecnum,rotating]{revtex}
\begin{document}

\def\cfp{{\it cfp }}
\def\2sp1lj{$^{2S+1}L_J$}
\def\fn{$f^N$}
\def\qfh{$4f^8$}
\def\qfs{$4f^6$}
\def\nln{$n\ell^N$}
\def\bkq{$B^k_q$}
\def\bzk{$B^k_0$}
\def\bzd{$B^2_0$}
\def\bqd{$B^2_q$}
\def\bzq{$B^4_0$}
\def\bqq{$B^4_q$}
\def\btq{$B^4_3$}
\def\bzs{$B^6_0$}
\def\bqs{$B^6_q$}
\def\bts{$B^6_3$}
\def\bss{$B^6_6$}
\def\stj#1#2#3#4#5#6{\left ( \begin{array}{ccc} #1 & #2 & #3 \\ #4 & #5 & #6 \end{array} \right )}
\def\ssj#1#2#3#4#5#6{\left \{ \begin{array}{ccc} #1 & #2 & #3 \\ #4 & #5 & #6 \end{array} \right \}}
\def\c#1{{\tt #1}}
\def\vol#1{{\bf #1}}
\def\cm{cm$^{-1}$}
\def\trait{\rule{5 cm}{.01in}}
\def\tbtio{Tb$_2$Ti$_2$O$_7$}
\def\eutio{Eu$_2$Ti$_2$O$_7$}
\def\eusno{Eu$_2$Sn$_2$O$_7$}
\def\ytio{(Tb$_{0.02}$Y$_{0.98}$)$_2$Ti$_2$O$_7$} 
\def\Eu{Eu$^{3+}$}
\def\Pr{Pr$^{3+}$}
\def\Nd{Nd$^{3+}$}
\def\tbt{Tb$^{3+}$}
\def\Tbq{Tb$^{4+}$}
\def\Er{Er$^{3+}$}
\def\Lu{Lu$^{3+}$}
\def\Y{Y$^{3+}$}
\def\Ln{Ln$^{3+}$}
\def\FJ{$^7$F$_J$}
\def\FO{$^7$F$_0$}
\def\FU{$^7$F$_1$}
\def\FD{$^7$F$_2$}
\def\FT{$^7$F$_3$}
\def\FQ{$^7$F$_4$}
\def\FC{$^7$F$_5$}
\def\FS{$^7$F$_6$}
\def\DJ{$^5$D$_J$}
\def\DO{$^5$D$_0$}
\def\DU{$^5$D$_1$}
\def\DD{$^5$D$_2$}
\def\DT{$^5$D$_3$}
\def\DQ{$^5$D$_4$}
\def\fdp{$4f^26p^1$}
\def\ft{$4f^3$}

\newcommand{\mubf}{\mbox{\boldmath $\mu$}}
\newcommand{\ti}{$\rm{Tb_{2}Ti_{2}O_{7}}$}
\newcommand{\tb}{$\rm{Tb^{3+}}$}
\newcommand{\Rij}{{\bf R}_{ij}({\bf n})}

\twocolumn[\hsize\textwidth\columnwidth\hsize\csname    
@twocolumnfalse\endcsname                               

\begin{title} {\LARGE \bf
Magnetic and Thermodynamic Properties of the \\
Collective Paramagnet$-$Spin Liquid
Pyrochlore
Tb$_2$Ti$_2$O$_7$}

\end{title} 

\author{
M. J. P. Gingras$^{1,2}$,
B. C. den Hertog$^2$,
M. Faucher$^3$,
J.S. Gardner$^{4,\dagger}$,
\\
L.J. Chang$^5$,
B.D. Gaulin$^{1,4}$,
N.P. Raju$^{6,\ddagger}$
J.E. Greedan$^6$
}

\address{$^1$Canadian Institute for Advanced Research}
\address{$^2$Department of Physics, University of Waterloo, Waterloo,
Ontario, N2L 3G1, Canada}
\address{$^3$Laboratoire SPMS, Ecole Centrale de Paris, 92295, Chatenay-Malabry, Cedex, France}
\address{$^4$Department of Physics and Astronomy, McMaster University,
Hamilton, Ontario, L8S 4M1, Canada}
\address{$^5$Institute of Physics, Academia Sinica, Taipei, Taiwan,
Republic of China}
\address{$^6$Brockhouse Institute for Materials Research and
Department of Chemistry, \\ McMaster University, Hamilton, Ontario, L8S
4M1, Canada}

\vspace{3mm}

\date{\today} 
\maketitle 

\begin{abstract}
{\noindent In a recent letter  [Phys. Rev. Lett. {\bf 82}, 1012 (1999)] it was 
found that the Tb$^{3+}$ magnetic moments in the Tb$_2$Ti$_2$O$_7$ 
pyrochlore lattice of corner-sharing tetrahedra remain in
a {\it collective paramagnetic} state down to
70mK.  In this paper we present results from d.c. magnetic susceptibility,
specific heat data, inelastic neutron scattering measurements, and crystal field 
calculations  that strongly suggest that (1) the Tb$^{3+}$ ions in 
Tb$_2$Ti$_2$O$_7$ possess a moment of approximatively 5$\mu_{\rm B}$, and (2) the
ground state $g-$tensor is extremely anisotropic below a temperature of $O(10^0)$K,  with Ising-like 
Tb$^{3+}$ magnetic moments confined to point along a 
local cubic $\langle 111 \rangle $ diagonal (e.g. towards the middle of the tetrahedron).
Such a very large easy-axis Ising like anisotropy along a $\langle 111 \rangle$
 direction dramatically reduces the frustration otherwise present in
a Heisenberg pyrochlore antiferromagnet. The results presented herein underpin the
conceptual difficulty in understanding the microscopic mechanism(s) responsible for
Tb$_2$Ti$_2$O$_7$ failing to develop long-range order at a temperature of the order 
of the paramagnetic Curie-Weiss temperature $\theta_{\rm CW} \approx -10^1$K.
We suggest that dipolar interactions and extra perturbative exchange coupling(s)
beyond nearest-neighbors may be responsible for the lack of ordering of 
Tb$_2$Ti$_2$O$_7$.  
}

\end{abstract}

\vskip2pc] 


\narrowtext

\section{Introduction}

\label{Intro}

\noindent In magnetic systems, competition between magnetic interactions,
combined with certain local lattice symmetries involving triangles, give
rise to the notion of {\it geometric frustration}~\cite{toulouse,reviews}.
Geometrically frustrated antiferromagnets are currently attracting much
interest within the condensed matter community~\cite{reviews,chandra}. The
main reason for this interest is that geometric frustration can cause
sufficiently large zero-temperature quantum spin fluctuations as to drive a
system into novel types of intrinsically quantum mechanical magnetic ground
states with no classical equivalent~\cite{chandra,lhuillier,lacroix}.

Among three-dimensional systems, the pyrochlore lattice of corner-sharing
tetrahedra (see Fig. 1)  with antiferromagnetic nearest-neighbor exchange interaction is
particularly interesting. For this system, theory~\cite
{villain,reimers-mft,moessner} and Monte Carlo simulations~\cite
{moessner,reimers-pyro} show that for classical Heisenberg magnetic moments
interacting with a nearest-neighbor antiferromagnetic coupling, there is no
transition to long-range magnetic order at finite temperature.
This is unlike the two-dimensional kagom\'e lattice 
antiferromagnet~\cite{moessner,chalker,reimers-kagome} where a thermally-driven
order-by-disorder of spin nematic order occurs. Villain
coined the name {\it ``collective paramagnetic''} to describe the 
{\it classical state} of the pyrochlore lattice at low temperatures~\cite{villain}.
\begin{figure}[h]
\begin{center}
  {
  \begin{turn}{0}%
    {\epsfig{file=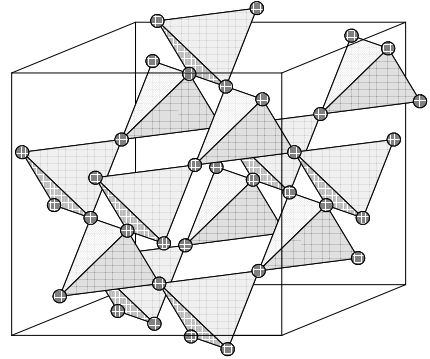,height=9.80cm,width=7cm} }
   \end{turn}
   }
\vspace{-35mm}
\caption{Pyrochlore lattice of corner-sharing tetrahedra. The magnetic moments occupy the
	corners of the tetrahedra.}
\end{center}
\end{figure}
\vspace{-5mm}
Because of their low propensity to order even for classical
spins, antiferromagnetic materials based on a pyrochlore lattice appear to
be excellent systems in which to seek exotic quantum mechanical ground
states. For example, numerical calculations suggest that the $S=1/2$
pyrochlore Heisenberg antiferromagnet may be fully quantum disordered,
giving rise to a state that is commonly referred to as 
{\it ``spin liquid''}~\cite{lacroix}.
 Both terms ``collective-paramagnet'' and
``spin liquid'' are meant to emphasize that despite such a system remaining
in a paramagnetic phase down to absolute zero temperature, the properties
of such a state involve very strong and nontrivial
short-range spin correlations, analogous to the nontrivial position-position
correlations present in an ordinary atomic or molecular fluid.

A number of experimental studies on insulating pyrochlore materials have
been reported in the past ten years. Interestingly, it has been found that
such systems do not typically form such a spin liquid state that remains
paramagnetic down to zero temperature. Most often, these systems
 either display
long-range antiferromagnetic order; such as FeF$_3$~\cite{fef3},
Gd$_2$Ti$_2$O$_7$~\cite{gd2ti2o7,gd2ti2o7-isis}, ZnFe$_2$O$_4$~\cite{znfe2o4,spinel},
and ZnCr$_2$O$_4$~\cite{spinel,zncr2o4}, or enter a spin-glass-like state
below some nonzero spin-freezing temperature as exhibited by 
Y$_2$Mo$_2$O$_7$~\cite{dunsiger,gingras_ymoo,gardner_y2mo2o7}, 
Tb$_2$Mo$_2$O$_7$~\cite{dunsiger,gaulin}, Y$_2$Mn$_2$O$_7$~\cite{y2mn2o7},
as well as the
disordered CsNiCrF$_6$ pyrochlore~\cite{csnicrf6}.

Recently, several studies of the pyrochlore rare-earth titanates,
R$_2$Ti$_2$O$_7$, have been published~\cite{re2ti2o7}. In these compounds, 
the trivalent
rare earth ions, R$^{3+}$, occupy the 16d sites of the Fd$\bar 3$m space and
form a pyrochlore lattice (Fig. 1). The behaviors displayed in this
family of pyrochlores are much varied indeed.
 Gd$_2$Ti$_2$O$_7$ develops true long-range order at a
critical temperature of about 1K~\cite{gd2ti2o7,gd2ti2o7-isis}. 
Tm$_2$Ti$_2$O$_7$ possesses a trivial non-magnetic (i.e. spin singlet) ground state
separated by an energy gap of about 120 K to the next crystal field 
level~\cite{tm2ti2o7}. Ho$_2$Ti$_2$O$_7$ is well described by an Ising 
doublet~\cite{ho2ti2o7}. In that system, it was originally argued that the
nearest-neighbor exchange interaction is weakly ferromagnetic~\cite{ho2ti2o7},
and that the strong Ising-like single ion anisotropy along  $(111)$
directions frustrates the development of long range ferromagnetic order~\cite
{ho2ti2o7,shastry,shastry-comm,bramwell-comm,hertog-ice,bramwell-ice,spin-ice,harris-comm}.
This material also exhibits low temperature spin dynamics reminiscent of
Pauling's ``ice model''~\cite{pauling}, an equivalence proposed by Harris
and co-workers~\cite{ho2ti2o7,bramwell-ice,spin-ice,harris-comm}. Recently,
it has been found that Dy$_2$Ti$_2$O$_7$~\cite{dy2ti2o7} is also a very good
example of ``spin ice''~\cite{ho2ti2o7,bramwell-ice,spin-ice,harris-comm},
and that the application of a magnetic field can restore much of the
ground-state entropy and drive magnetic phase transitions. Most recently,
den Hertog and Gingras have argued that the spin ice physics in both Ho$_2$Ti$_2$O$_7$
and Dy$_2$Ti$_2$O$_7$ is not driven by nearest-neighbor
ferromagnetic exchange, but is rather due to the {\it long-range} $1/r^3$
nature of magnetic dipole-dipole interactions~\cite{bramwell-comm,hertog-ice}.

In contrast to the long-range ordered or spin glass states mentioned
above, strong evidence for collective paramagnetism, or spin liquid
behavior, was recently observed in the insulating pyrochlore Tb$_2$Ti$_2$O$_7
$~\cite{tb2ti2o7,tb2ti2o7-field}. It was found using neutron scattering and
muon spin relaxation methods that this material remains paramagnetic down to
(at least) 70 mK despite the fact that the paramagnetic Curie-Weiss temperature, 
$\theta_{{\rm CW}}$, is -19 K, and that short-range antiferromagnetic
correlations begin to develop at $\sim$ 50 K. At first sight, one could argue that
it is ``pleasing'' to have found at last the spin liquid state anticipated
by theory for a highly frustrated pyrochlore antiferromagnet~\cite
{villain,reimers-mft,moessner,reimers-pyro}. However, the situation for Tb$_2$Ti$_2$O$_7$
is not
as simple as it might naively appear.

In Tb$_2$Ti$_2$O$_7$
the Tb$^{3+}$ ions have
a partially filled $^7F_6$ shell, and one most first understand
their crystal field level scheme and, in particular, the nature of the
single-ion magnetic ground state before constructing a correct effective
spin-spin Hamiltonian for Tb$_2$Ti$_2$O$_7$. Indeed, we show below that 
crystal field anisotropy renders the description of Tb$_2$Ti$_2$O$_7$ in
terms of an isotropic Heisenberg antiferromagnetic model completely
inappropriate.

If one neglects the axial oxygen distortion around the Tb$^{3+}$ sites,
and assumes that the local environment of the Tb$^{3+}$ is
perfectly cubic, one would expect, based on point charge calculations, that
the ground state of both Tb$^{3+}$ and Tm$^{3+}$ should either be a singlet
or a nonmagnetic doublet~\cite{lea}. For example, as mentioned above,
 experimental evidence 
for a nonmagnetic
singlet ground state has been found in Tm$_2$Ti$_2$O$_7$~\cite{tm2ti2o7}.
Based on this naive picture,
one can see that the
experimental evidence of a moment for Tb$^{3+}$ in Tb$_2$Ti$_2$O$_7$
is therefore a nontrivial issue that needs to be understood.

A simple possibility is that corrections beyond the point-charge approximation and/or
the known axial oxygen distortions around each of the 16d sites cause the
Tb$^{3+}$ cations to acquire a permanent magnetic moment. Another and more
interesting possibility,
is that the moment on the Tb$^{3+}$ site in Tb$_2$Ti$_2$O$_7$ is induced
by a collective bootstrapping of the magnetic (exchange and/or dipolar)
interactions as occurs in the tetragonal LiTbF$_4$ material~\cite{litbf4}.
In Tb$_2$Ti$_2$O$_7$, a priori, it is theoretically possible that there could be no
moment on the Tb$^{3+}$ site for a concentration $x$ of Tb$^{3+}$ less than some
critical concentration $x_c$ in (Tb$_x$Y$_{1-x}$)$_2$Ti$_2$O$_7$,
as occurs in LiTb$_x$Y$_{1-x}$F$_4$~\cite{litbf4}. This is an important
issue. Indeed, one could imagine that for the highly-frustrated pyrochlore lattice,
the collective development of a permanent ground-state moment
would not give rise to homogeneous moments on the Tb$^{3+}$ sites, but to a
kind of ``modulated moment structure''. This idea is conceptually similar to
what is found in the frustrated tetragonal TbRu$_2$Ge$_2$ material~\cite{charnier},
but where for Tb$_2$Ti$_2$O$_7$ there might be instead
a quantum-disordered state ``intervening'' between a trivial singlet ground
state and a long-range ordered one, with the quantum-disordered state
extending all the way to $x=1$.  

In
the case where a permanent moment does exist on Tb$^{3+}$ even in absence of
interaction (i.e. the limit $x\rightarrow 0$ in (Tb$_x$Y$_{1-x}$)$_2$Ti$_2$O$_7$),
the important issue is to determine the wavefunction decomposition of
the ground state in terms of $\vert J,M_J\rangle$ states and the symmetry,
Heisenberg or otherwise,  of the resulting effective spin variable. The
goal of such a programme is to construct a low-energy effective spin
Hamiltonian in
order to tackle theoretically why Tb$_2$Ti$_2$O$_7$ does not order at
nonzero temperature. Consequently, it is very important to know in more
details what the magnetic nature of the Tb$^{3+}$ single-ion ground state in
Tb$_2$Ti$_2$O$_7$ is.

The main purpose of this paper is to examine the magnetic nature of the Tb$^{3+}$
ion in the Tb$_2$Ti$_2$O$_7$ pyrochlore in order to assess whether or not
there is indeed a {\it permanent} moment at the Tb site as the temperature
goes to zero, and determine the nature of this moment (e.g. level of
effective spin anisotropy). We present in Section II experimental evidence,
based on results from d.c. susceptibility, heat capacity and powder
inelastic neutron studies that show there is a permanent moment at the Tb
site, but that its approximate 5$\mu_{{\rm B}}$ value is less than the value
of 9.4$\mu_{{\rm B}}$ estimated from the d.c. susceptibility measurements
above 200K~\cite{tb2ti2o7}, or the 9.72$\mu_{{\rm B}}$ $^7F_6$ free ion
value. To complement the experimental work, results from ab-initio crystal
field calculations that take into account covalent and electrostatic effects
are presented in Section III and Appendix B. We discuss in Section IV the possibility that
dipole-dipole interactions and extra perturbative exchange couplings beyond nearest-neighbor
may be responsible for the lack of ordering in Tb$_2$Ti$_2$O$_7$.


\section{Experimental Method \& Results}

\subsection{Sample Preparation}

Samples of Tb$_2$Ti$_2$O$_7$ and (Tb$_{0.02}$Y$_{0.98}$)$_2$Ti$_2$O$_7$ were
prepared in the form of polycrystalline pellets by high temperature solid
state reaction. Starting materials, Tb$_2$O$_3$, Y$_2$O$_3$ and TiO$_2$ were
taken in stoichiometric proportions, mixed thoroughly, pressed into pellets
and heated in an alumina crucible at 1400$^o$C for 12 hours in air. Tb$_2$O$_3$
was prepared by hydrogen reduction of Tb$_4$O$_7$. The powder x-ray
diffraction patterns of the samples, obtained with a Guinier-Hagg camera,
indicate that they are single phase with cubic unit cell constants, a$_0$,
of 10.491$\AA$ for Tb$_2$Ti$_2$O$_7$ and 10.104$\AA$ for 
 (Tb$_{0.02}$Y$_{0.98}$)$_2$Ti$_2$O$_7$. The value for the concentrated sample is in
excellent agreement with previous reports~\cite{brixner,knop}.

Some of this high quality
polycrystalline material was then used as starting material for
a successful single crystal growth using an
optical floating zone image furnace.  Details of
the crystal growth are described elsewhere~\cite{xtal}.


\subsection{d.c. Magnetic Susceptibility Measurements}

As a first step towards determining the magnetic nature of the electronic ground
state of the Tb$^{3+}$ cations in Tb$_{2}$Ti$_{2}$O$_{7}$, we have
investigated the d.c. magnetic susceptibility of Tb$_{2}$Ti$_{2}$O$_{7}$ and
(Tb$_{0.02}$Y$_{0.98}$)$_{2}$Ti$_{2}$O$_7$.
The d.c. magnetic susceptibility was measured using a SQUID magnetometer(Quantum
Design, San Diego) in the temperature range 2-300 K. The inverse
susceptibility, $\chi ^{-1}$, of Tb$_{2}$Ti$_{2}$O$_{7}$ measured at an
applied field of 0.01 Tesla is shown in Fig. 2. A fit of the data to the
Curie-Weiss (CW) law above 200 K gives an effective paramagnetic moment of
9.4$\mu _{{\rm B}}$/Tb$^{3+}$ and an effective (high-temperature) CW
temperature, $\theta _{{\rm CW}}$ = -19 K, which indicates the dominance of
antiferromagnetic interactions (the Curie-Weiss fits are done above 200 K).
Deviation from the Curie-Weiss law sets in at a rather high temperature, 
$T\sim $70 K. This is consistent with elastic neutron scattering results
reported previously, which showed evidence for short range magnetic
correlations at temperatures up to 50 K~\cite{tb2ti2o7}.

Recognizing that for Tb$^{3+}$, an $^{7}$F$_{6}$ even electron and
non-S-state ion, there may be ``crystal field'' as well as exchange
contributions to the experimentally determined 
$\theta _{{\rm CW}}$, a magnetically dilute sample, 
(Tb$_{0.02}$Y$_{0.98}$)$_{2}$Ti$_{2}$O$_7$,
was also studied. The data for this sample are shown in Fig. 3 along
with a Curie-Weiss fit giving again a value close to the free ion value
for the effective paramagnetic moment, 9.6$\mu _{{\rm B}}$/Tb$^{3+}$, and 
$\theta _{{\rm CW}}\approx -6$ K. This finite value is in contrast to the
essentially zero $\theta _{{\rm CW}}$ value obtained for a similarly diluted
sample of Gd$_{2}$Ti$_{2}$O$_{7}$ which contains the isotropic ``spin only'' 
$^{8}$S$_{7/2}$ Gd$^{3+}$ ion \cite{gd2ti2o7}, and therefore indicates that
a significant crystal field contribution to $\theta_{\rm CW}$
exists in the Tb$^{3+}-$based
material. Thus, to a first approximation one can estimate that the portion
of $\theta _{{\rm CW}}$ for Tb$_{2}$Ti$_{2}$O$_{7}$ which can be attributed
to magnetic {\it interactions} is \newline 
$\theta_{\rm CW}$$\{$Tb$_{2}$Ti$_{2}$O$_{7}$ $\}$
-
$\theta_{\rm CW}$$\{$(Tb$_{0.02}$Y$_{0.98}$)$_{2}$Ti$_{2}$O$_7$ $\}$
$\sim $ -13 K, \newline
a value similar to the 
$\theta _{{\rm CW}}\approx $ -10 K  found for Gd$_{2}$Ti$_{2}$O$_{7}$~\cite{gd2ti2o7}.
This approach can be made more rigorous by noting that in a
high-temperature series expansion, one finds that the magnetic
susceptibility is $\chi =C_{1}(1/T+C_{2}/T^{2})$ where $C_{2}\equiv\theta _{{\rm CW}}
$ can be ``decomposed'' as a simple sum of terms that are ascribed to
exchange interactions, dipolar interactions and crystal-field terms (see
Appendix A). Note also that, down to a temperature of $T=2$ K, neither Fig.
2 nor Fig. 3 show any sign of a singlet ground state which would manifest
itself as an approach to a constant susceptibility with decreasing
temperature, as found in the single ground-state of the Tm$_{2}$Ti$_{2}$O$_{7}$~\cite{tm2ti2o7}.
A more detailed analysis concerning this issue is presented in
Section III.

While it is tempting to use the magnetic interaction contribution 
($\theta_{{\rm CW}}-\theta _{{\rm CW}}^{{\rm cf}})\approx -13$ K,
(with $\theta_{\rm CW}\approx -19$ K and
$\theta_{\rm CW}^{\rm cf} \approx -6$ K), to extract the
approximate value of the nearest neighbor exchange, an estimate of the
nearest neighbor dipole-dipole interaction for ${\rm Tb^{3+}}$ ions
indicates an energy scale of about 1 K. Given the long range nature of
dipolar forces, it becomes clear that the classical nearest neighbor
exchange constant cannot be obtained from a measurement of the Curie-Weiss
temperature until the effects of long range dipolar interactions on 
$\theta_{\rm CW}$
have been understood.  We find via a high temperature series
expansion analysis of the long range dipolar contributions,
$\theta _{{\rm CW}}^{{\rm dip}}$, to the Curie-Weiss temperature (see
Appendix A), that the
estimated upper bound on $\theta _{{\rm CW}}^{{\rm dip}}$ is ferromagnetic
and $\sim 1.2$ K (for needle-shaped powder crystallites), while the lower
bound is antiferromagnetic and $\sim -2.4$ K (for slab-shaped powder
crystallites). Consequently, we find that antiferromagnetic exchange
interactions are
predominantly responsible for the 
($\theta _{{\rm CW}}-\theta _{{\rm CW}}^{{\rm cf}})=-13$ K value determined above $T\gtrsim 200$ K,
with a resulting $\theta_{\rm CW}^{\rm exchange} \in [-14.2,-10.6]$ K.
\begin{figure}
\begin{center}
  {
  \begin{turn}{0}%
    {\epsfig{file=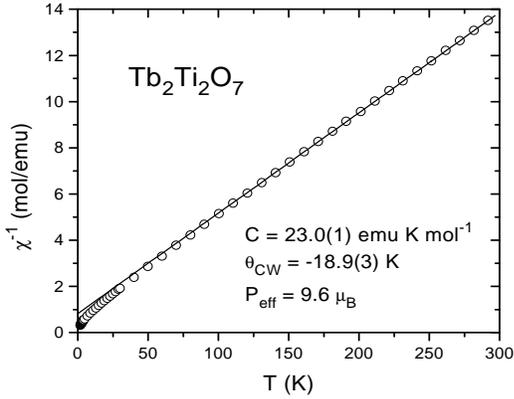,height=10cm,width=8cm} }
   \end{turn}
   }
\vspace{-2cm}
\caption{Inverse molar susceptibility, $1/\chi$, of Tb$_2$Ti$_2$O$_7$
vs temperature.}
\end{center}
\end{figure}
\vspace{-1cm}
\begin{figure}
\begin{center}
  {
  \begin{turn}{0}%
    {\epsfig{file=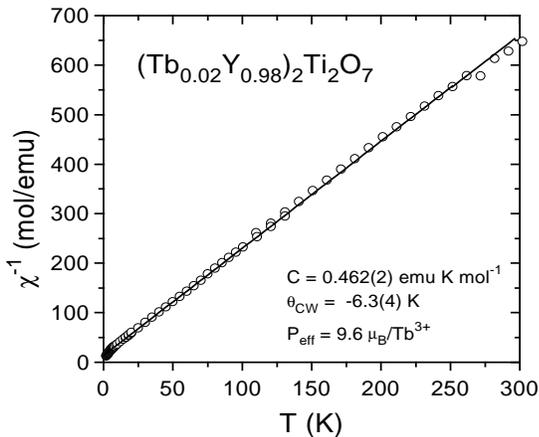,height=10cm,width=8cm} }
   \end{turn}
   }
\vspace{-2cm}
\caption{Inverse molar susceptibility, $1/\chi$,  of the diluted
(Tb$_{0.02}$Y$_{0.98}$)$_2$Ti$_2$O$_7$ material vs
temperature.}
\end{center}
\end{figure}


\subsection{Neutron Scattering Experiments}

Inelastic neutron scattering allows us to determine with reasonable precision the
values of the electronic energy levels of Tb$_{2}$Ti$_{2}$O$_{7}$.
 The inelastic neutron scattering
measurements were carried out on a 50g sample of polycrystalline 
Tb$_{2}$Ti$_{2}$O$_{7}$, loaded in a sealed Al cell with a helium exchange gas present.
The cell was mounted in a closed cycle helium refrigerator with a base
temperature of 12 K. The sample was the same as used in reference~\cite
{tb2ti2o7}. Measurements were performed on the C5 triple axis spectrometer
at the Chalk River Laboratories in constant scattered neutron energy mode.
Two spectrometer configurations were employed, appropriate for relatively
high and low energy resolution, respectively. Both configurations employed
pyrolitic graphite (PG) as both monochromator and analyser. The low
resolution measurements, appropriate for relatively high energy transfers,
were performed using ${\frac{E^{\prime }}{h}}$=3.52 THz
(1THz=48 K), open-60-80-open
collimation, and a PG filter in the scattered beam. The high resolution
configuration used ${\frac{E^{\prime }}{h}}$=1.2 THz, open-40-60-open
collimation, and a cooled Be filter in the scattered beam. 
These results clearly indicate excitations at $E\approx $ 0.35THz, 2.5THz
and a broad neutron group centered at 3.5THz (corrsponding to
16.8 K, 120 K, and 168 K, respectively).

The low energy-resolution inelastic neutron scattering measurements at 12 K
with ${\frac{E^{\prime}}{h}}$=3.52 THz revealed the presence of two 
$Q$-independent modes with frequencies $\nu \sim$ 2.5 and 3.5 THz.
Representative neutron groups, as well as the dispersion relation for these
two excitations are shown in the top panel of Fig. 4. These excitations are
identified as being magnetic in origin due to their temperature and Q
dependence. Their flat dispersion indicates that they are crystal electric
field levels for Tb$^{3+}$ in the environment appropriate for 
Tb$_2$Ti$_2$O$_7$. 

The high energy-resolution inelastic neutron measurements at 12 K with
${\frac{E^{\prime}}{h}}$=1.2 THz shows the presence of a low lying magnetic
excitation near $\nu\sim$ 0.35 THz. This mode is also dispersionless above a
temperature of $\sim$ 25 K, but partially softens in energy at the
wavevector which characterizes the very short range spin correlations, which
develop below 25 K. The development of this interesting dispersion has been
described previously~\cite{tb2ti2o7}. The dispersion of the low lying
excitation is shown in the lower panel of Fig. 4. One can see that this
partial softening of the excitation branch occurs only for the lowest lying
mode. As we show in the next section, this is manifested
in the heat capacity measurements as broad features that 
result of a broadening of the single-ion energy levels via these magnetic
correlation effects.

These measurements place constraints on any calculations for the energy
eigenstates of Tb$^{3+}$ as they set both the energy spacing of the levels,
and require that magnetic dipole matrix elements must connect the ground
state with these levels in order that they be visible to the inelastic
neutron scattering experiment. In particular, this indicates nonzero
$\langle 0 \vert J^+ \vert 1 \rangle$, $\langle 0 \vert J^- \vert 1 \rangle$,
or $\langle 0 \vert J^z \vert 1 \rangle$ matrix elements connecting the
ground state, $\vert 0\rangle$, and the first excited state, $\vert 1\rangle$,
at an energy $\sim 0.35$ THz $\approx$ 17 K above the ground state. In
other words, there must be large $\vert J, M_J\rangle$ components in $\vert
0\rangle$ and $\vert 1\rangle$ where some of the $M_J$ involved for the
ground state and the excited state differ by 0, $\pm 1$.
\begin{figure}
\begin{center}
  {
  \begin{turn}{0}%
    {\epsfig{file=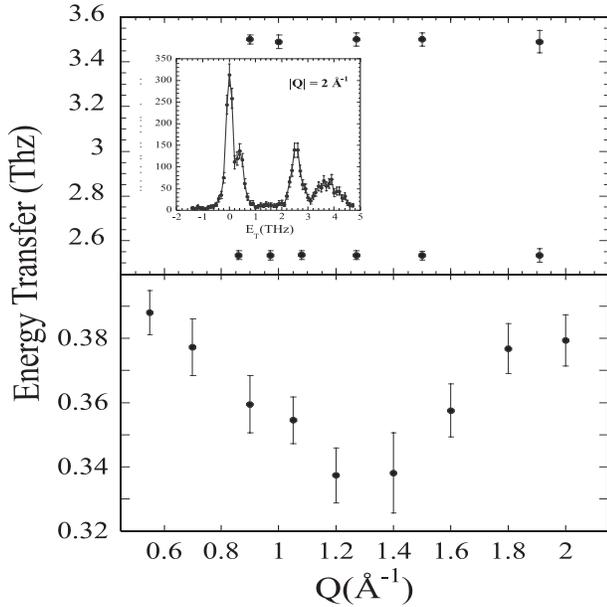,height=8cm,width=8cm} }
   \end{turn}
   }
\caption{
Inelastic neutron scattering data obtained at a temperature
$T=12$ K.
The inset of the top panel of
Fig. 4 shows a transfer energy scan at $Q=2\AA^{-1}$.
Data showing representative neutron
groups as well as the dispersion relation for these excitations at $\sim$ 2.5
and 3.5 THz
is shown in the top panel of Fig. 4.
The dispersion of the low lying excitation
is shown in the lower panel.}
\end{center}
\end{figure}



\subsection{Specific Heat Measurements}

Low-temperature specific-heat measurements on Tb$_2$Ti$_2$O$_7$ were
performed using a thermal-relaxation microcalorimeter. The single crystal
sample was mounted on a sapphire holder which was isolated from the bath by
four copper-gold alloy wires. The relative precision and absolute accuracy
of the calorimeter were confirmed by measuring copper and gold standards. In
principle, specific heat-measurements on a dilute
(Tb$_{0.02}$Y$_{0.98}$)$_2$Ti$_2$O$_7$ sample would be 
useful.
Unfortunately, this is not technically easily feasible as the magnetic
contribution to the total specific heat would be too small to be determined
accurately.
\begin{figure}
\begin{center}
  {
  \begin{turn}{0}%
    {\epsfig{file=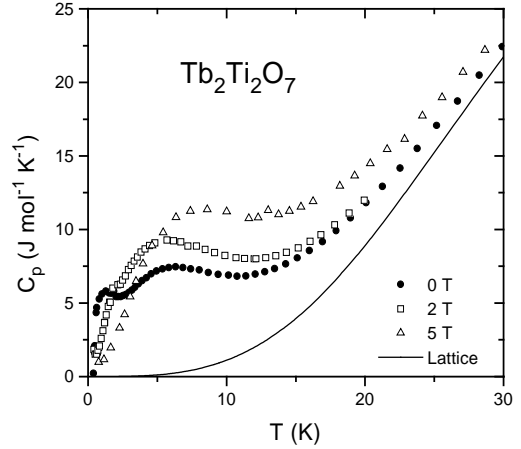,height=11cm,width=8cm} }
   \end{turn}
   }
\vspace{-2cm}
\caption{Specific heat, $C_p$, of  Tb$_2$Ti$_2$O$_7$
as function of temperature for the indicated magnetic fields of 0, 2 and 5 T. The solid
line corresponds to the lattice specific heat of  Tb$_2$Ti$_2$O$_7$,
$C_l$, estimated from
the measurements on the non-magnetic Y$_2$Ti$_2$O$_7$ that is isotructural to
 Tb$_2$Ti$_2$O$_7$ [52].}
\end{center}
\end{figure}
The total specific heat, $C_p$, of Tb$_2$Ti$_2$O$_7$ was measured from 0.4 K
to 30 K at applied fields of 0, 2 and 5 T (see Fig. 5). The zero field data
exhibits two broad peaks centered at about 1.5 K and 6 K. The data are in
agreement with those of Ref.~\cite{shastry}, above a temperature of 0.4 K.
Hyperfine contributions to the specific heat become important
below 0.4 K for Tb$-$based compounds, as found for example in the 
Tb$_2$(GaSn)O$_7$ pyrochlore~\cite{tb2ga2sn2o7,tb-hyperfine}, and this is 
presumably the reason for the sharp increase of $C_p(T)/T$ found in Fig. 4 of Ref.~\cite{shastry}
below 0.5 K. The solid line in Fig. 5 corresponds to the estimated lattice
heat capacity, $C_l$, for Tb$_2$Ti$_2$O$_7$, determined by scaling the heat
capacity for Y$_2$Ti$_2$O$_7$, which is insulating, non-magnetic, and is
isotructural to Tb$_2$Ti$_2$O$_7$~\cite{gmelin}.

The magnetic specific heat, $C_{m}$, obtained by subtracting $C_{l}$ from 
$C_{p}$, is shown in \mbox{Fig. 6} for the three applied fields. With the
application of a 2 T field the magnitude of the lower temperature peak
diminishes and moves to a slightly higher temperature. In contrast, the
position of the second peak does not change but increases in magnitude as
the lower temperature feature begins to overlap with it. At 5 T the low
temperature feature disappears completely and the remaining peak is shifted
to higher temperatures. 
\begin{figure}
\begin{center}
  {
  \begin{turn}{0}%
    {\epsfig{file=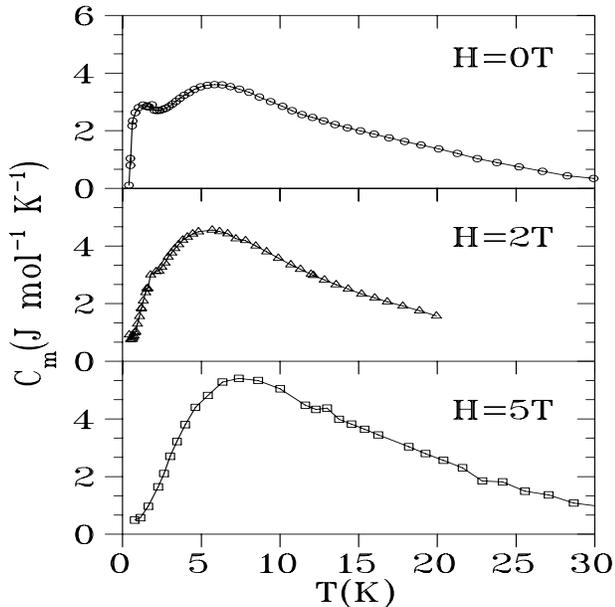,height=8cm,width=8cm} }
   \end{turn}
   }
\caption{Magnetic specific heat, $C_m$, of  Tb$_2$Ti$_2$O$_7$
as function of temperature for the indicated magnetic fields of 0, 2 and 5 T.
The solid lines are guides to the eye.}
\end{center}
\end{figure}
The crystal field calculations described in Section III and Appendix B (see
Table 2) indicate a level scheme consisting of a ground state doublet with
another doublet as the first excited state. Attempts to fit the $C_{m}$ data
to Schottky anomalies using a unique doublet$-$doublet level scheme for the
Tb$^{3+}$ ions and varying the ground state$-$excited state energy splitting
failed, since important magnetic short-range correlations are present in 
this sytem, as discussed above in neutron scattering~\cite{tb2ti2o7}.
 Consequently, we interpret the anomaly at $\sim$ 6 K as a remnant of the excited doublet
that is ``broadened'' by exchange correlation fields, while 
the 1.5K anomaly is
presumably due to these same correlation effects, but now acting on the
single-ion ground state doublet. In other words, the build-up of
short-range correlations in the low-temperature sector of the effective
Hamiltonian for Tb$_{2}$Ti$_{2}$O$_{7}$ results in a specific heat anomaly
at 1.5 K.  This low-temperature anomaly at 1.5K that results from
correlations is akin to the broad specific heat bump at $\approx$ 2 K
found in Gd$_{2}$Ti$_{2}$O$_{7}$~\cite{gd2ti2o7}. 
 However, Gd$_{2}$Ti$_{2}$O$_{7}$ is an $^{8}S_{7/2}$ 
spin-only ion, and there are no crystal field levels at high energy, 
nor are there correlation remnants of crystal field levels above the
ground state such as those
that cause the specific heat anomaly at 6\ K in Tb$_{2}$Ti$_{2}$O$_{7}$.
Another distinction between Tb$_{2}$Ti$_{2}$O$_{7}$ and 
Gd$_{2}$Ti$_{2}$O$_{7}$ is that the latter shows a very 
sharp specific heat anomaly at 0.9 K~\cite{gd2ti2o7}, 
and therefore presumably a transition to long range order
at that temperature as suggested by recent neutron scattering experiments
\cite{gd2ti2o7-isis}. Down to 0.4 K, no such sharp specific heat anomaly is
found in Tb$_{2}$Ti$_{2}$O$_{7}$.  From the fit of the d.c.
susceptibility and the crystal field calculations presented in Section III
and Appendix B,
good evidence is obtained that the magnetic moment in the ground state and
the first excited state is $\sim $ 5 $\mu _{\text{B}}$ and $\sim 6$ $\mu _{\text{B}}$,
respectively. Given a doublet-doublet energy gap of about 17 K, we can
estimate the strength of the magnetic field where the separation between the
doublets is equal to the magnetic field energy, and find a magnetic field of
about 5 T. For an applied field of that strength, the ground state and
excited states merge and are strongly mixed. This explains the
disappearance
 of the low-temperature specific heat anomaly in Fig. 6 for a fuekd $H=$ 5 T.

It is usual to estimate the magnetic entropy, $S_m(T)$,  in a system by
integrating $C_m(T)/T$ between the lowest temperature reached
and the temperature, $T$, of interest. Although it is straightforward
to integrate $C_m(T)/T$, the interpretation of the results
for Tb$_2$Ti$_2$O$_7$ is difficult. The main reasons for this are:
\begin{itemize}
\item  The hyperfine interaction is large for Tb$^{3+}$  and the nuclear
          specific heat contribution becomes significant with respect to
          the magnetic contribution near 0.4K~\cite{tb2ga2sn2o7,tb-hyperfine}.
\item There is a doublet crystal field excitation at an energy of approximately 17 K.
      Hence, one can hardly integrate $C_m(T)/T$ above 5 K without ``already''
	 embedding in the resulting entropy a contribution from
      excitations to the first excited doublet.
\item From our neutron results and crystal field calculations
presented in Appendix B, it is known that there are other
levels at an energy $\sim 10^2$ K. 
Hence one cannot 
          integrate $C_m(T)/T$ to obtain $S_m(T)$  up to a high enough
          temperature without having to consider the specific heat
          contribution from the states at $\sim O(10^2)$ K.
\item  By integrating $C_m(T)/T$ up
         to $\approx 30$ K 
	one enters a regime
          where the lattice contribution to the specific heat, $C_l(T)$,
          becomes sizeable (see Fig. 5). In that case the subtraction of
          $C_l(T)$ from the total $C_p(T)$ using rescaled results for the 
          isostructural non-magnetic Y$_2$Ti$_2$O$_7$ leads to inherent
          uncertainties which increase dramatically above 10 K (see Fig.
          5).
\end{itemize}
With these provisions in mind, 
we have determined 
$S_m(T)$ (see Fig. 7).
 We find that the recovered entropy, $S_m(T)$,  at 15 K is already
larger than that expected for a singlet$-$doublet energy level
scheme, $S_m(15 {\rm K})> {\rm R} \ln(3)$. 
Since 15 K is much less than the excited states at $\sim 100$ K, there should
be little contribution to
$S_m(15 {\rm K})$
coming from states at $T\gtrsim$ 100 K.
Consequently, the results presented in this figure support further
the picture that the two lowest energy levels in 
Tb$_2$Ti$_2$O$_7$ consists of two doublets.  However, it 
is interesting to
note that at a temperature of 30 K, the recovered entropy is not yet
equal to R$\ln(4)$, the total entropy for two doublets. Either
30 K is not yet at high enough temperature to have recovered the
full doublet$-$doublet entropy, or there exists macroscopic entropy
in the ground state as occurs in Dy$_2$Ti$_2$O$_7$~\cite{dy2ti2o7}.
For the four reasons mentioned above, it is difficult to make the
discussion about the recovered entropy above 30 K in
Tb$_2$Ti$_2$O$_7$ much more quantitative.

In summary the magnetic specific heat data are consistent with
the inelastic neutron scattering results in
that the two lowest-lying energy levels for Tb$^{3+}$ in Tb$_{2}$Ti$_{2}$O$_7$
consist of two doublet energy levels separated by an excitation energy
of $\sim $ 15 $-$ 20 K.
As we will show in the next Section and Appendix B, our crystal field calculations 
strongly suggest that these two lowest lying levels are
magnetic (Ising) doublets.
\begin{figure}
\begin{center}
  {
  \begin{turn}{90}%
    {\epsfig{file=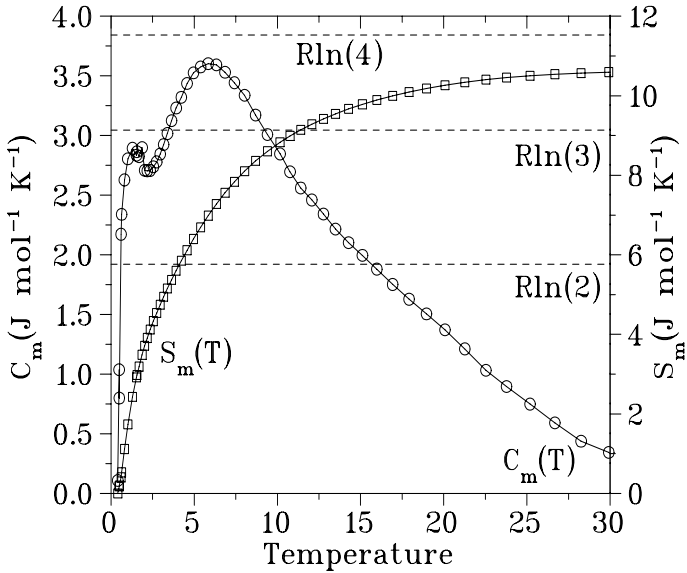,height=9cm,width=7.5cm} }
   \end{turn}
   }
\caption{Magnetic specific heat, $C_m(T)$, of  Tb$_2$Ti$_2$O$_7$ in
zero applied field (circles), and entropy, $S_m(T)$ (squares) are shown.
The entropy values for two, three and four states are indicated by
the horizontal dashed lines. R is the molar gas constant R$=N_0k_{\rm B}$
=8.3145 J mol$^{-1}$ K$^{-1}$.
The solid lines are guide to the eye.}
\end{center}
\end{figure}


\section{Single ion Properties}

As discussed in the introduction, it is very important to determine whether the 
existence of a moment at the Tb$^{3+}$ site in Tb$_2$Ti$_2$O$_7$ is intrinsic,
or driven by magnetic interactions. Consequently, we have investigated in detail
the problem of single-ion properties of Tb$^{3+}$, both theoretically and via
d.c. susceptibility measurements, of
Tb$_2$Ti$_2$O$_7$  and of the dilute
(Tb$_{0.02}$Y$_{0.98}$)$_2$Ti$_2$O$_7$
material where the Tb$-$Tb interactions should be negligible.
The main conclusion from this investigation is that the
Tb$^{3+}$ cation does indeed carry an intrinsic moment inherent to its
environment in Tb$_2$Ti$_2$O$_7$, and that moment is not due to
a bootstrapping effect from interactions.

The results from inelastic neutron scattering and specific heat measurements
presented in the previous section already provide some evidence for a doublet
ground state and an excited doublet state at an energy of $\sim$ 17K. The purpose of this
section is to investigate further from a theoretical
point of view the question of the existence of a magnetic
doublet ground state for Tb$^{3+}$ in Tb$_2$Ti$_2$O$_7$.
In short, both a simple point
charge calculation and a more 
sophisticated ab-initio method
confirm that a doublet$-$doublet scheme
is the most likely low energy level structure for Tb$^{3+}$.
We have constructed a van Vleck equation based on such a doublet$-$doublet
scheme in order to parametrize
the d.c. suceptibility of
the dilute
(Tb$_{0.02}$Y$_{0.98}$)$_2$Ti$_2$O$_7$ sample, and
to extract the value of the magnetic moment for both the ground state and
excited state doublets. We find that experimental results for the 
d.c. susceptibility of (Tb$_{0.02}$Y$_{0.98}$)$_2$Ti$_2$O$_7$ are
in reasonably good agreement with the ab-initio calculations.

\subsection{Crystal Field Effects}

Experimental evidence of a doublet-doublet structure for the low 
temperature crystal field 
levels of ${\rm Tb^{3+}}$ in Tb$_2$Ti$_2$O$_7$
can be understood by considering the 
crystal field environment surrounding the ion.
Pyrochlore oxides $\rm{A_{2}B_{2}O_{7}}$  are described in space group  Fd$\bar 3$m with A$^{3+}$, the 
    trivalent rare earth in 16d, B$^{4+}$, the tetravalent transition metal ion,
    in 16c, O1 in 48e and O2 in 8b. The A or Tb$^{3+}$ site in this case
    is coordinated to six O1 ions at about 2.5$\AA$  in the form of a puckered
    ring and to two O2 ions at a distance of 2.2$\AA$ in the form of a linear
    O2-Tb-O2 chain oriented normal to the mean plane of the O1 ring.
    The O2-Tb-O2 units are parallel to the $(111)$ directions  within the 
    cubic unit cell. Overall, the local geometry at the Tb$^{3+}$ site
    can be described as a severe trigonal compression along the body 
    diagonal of a simple cube.

Based on symmetry considerations, the cubic plus axial distortion   
surrounding  the $\rm{Tb^{3+}}$ ion  may be expressed by a 
crystal field
Hamiltonian of the general form
\begin{eqnarray}
\label{cef}
H^{\rm cf}&=&B^{2}_{0}C^{2}_{0} + B^{4}_{0}C^{4}_{0} + B^{4}_{3}C^{4}_{3} 
\nonumber \\
&+&B^{6}_{0}C^{6}_{0} + B^{6}_{3}C^{6}_{3} + B^{6}_{6}C^{6}_{6} \;,  
\end{eqnarray}
where the $B^{k}_{q}$'s are yet to be determined crystal field
parameters, and the  $C^{k}_{q}$'s are tensorial operators defined as
$C^{k}_{q}=[4\pi/(2k+1)]^{1/2}Y^{q}_{k}$, where $Y^{k}_{q}$ is a
normalized spherical harmonic.  In general, the $B^{k}_{q}$'s represent an
effective one-body potential which lifts the degeneracy of the 
angular momentum states in question. In practice, they may be determined 
experimentally by spectroscopic or thermodynamic probes, or theoretically 
within various levels of approximation. 

In the simplest approximation of the crystal field interactions, one often uses the
so-called point charge (PC) approximation where the crystal field is simply assumed to
be caused by the
Coulomb field of point charges situated at neighboring sites. 
In such a picture, the 
crystal field eigenstates and eigenvalues 
have been determined for a number of
rare earths by Lea {\it et al.} for systems with
cubic symmetry (i.e. without trigonal distortion)~\cite{lea}. 
For $\rm{Tb^{3+}}$, the 
lowest three energy levels   in a cubic environment
are a $\Gamma_{3}$ singlet, a non-magnetic $\Gamma_{2}$ doublet and a
$\Gamma^{(2)}_{5}$ triplet, with their precise ordering in terms of 
lowest energy states dependent upon variation of  the point charge crystal field parameters 
\cite{lea}.  In general, the addition of the  trigonal
distortion splits the $\Gamma_{5}^{(2)}$ triplet into a singlet and doublet, 
while the $\Gamma_{3}$ and $\Gamma_{2}$ states are preserved. Hence, we expect that the 
crystal field
ground state of  the ${\rm Tb^{3+}}$ ion in Tb$_2$Ti$_2$O$_7$
will be a competition between two doublets and two singlets.

Some notion regarding the difficulty of determining the precise ordering of these
states and the size of their associated magnetic moments  may be obtained by 
diagonalizing  the crystal field
Hamiltonian $H^{\rm cf}$ using Stevens' operator equivalents \cite{Stevens} of the
$C_{q}^{k}$ within a fixed $J$ manifold\cite{Kassman},  and by using the
point charge approximation for the Coulomb effects of  the surrounding
oxygen ions. The resulting crystal field point charge
Hamiltonian, H$_{\rm pc}^{\rm cf}$,  with   quantization axis along the appropriate  
 $\langle 111 \rangle$
direction 
can be expressed as,
\begin{eqnarray}
\label{pc}
H^{\rm cf}_{\rm pc}&=&\alpha_{J}\tilde B^{0}_{2}(r^{3} -1)O^{0}_{2} \nonumber \\
&+& \beta_{J}\tilde B^{
0}_{4}\left[ \frac{(27r^{5} + 1)}{28}O^{0}_{4} -
20\sqrt{2}O^{3}_{4}\right] \nonumber \\ 
&+&\gamma_{J}\tilde B^{0}_{6}\left[\frac{(188 + 324r^{7})}{512}O^{0}_{6} +
\frac{35\sqrt 2}{4}O^{3}_{6} + \frac{77}{8}O^{6}_{6}\right] ,
\end{eqnarray}
where $r=R_{1}/R_{2}$ and $R_{1},R_{2}$ are the Tb-O distances for oxygen
ions situated 
on the puckered ring and on the distortion axis respectively. The
$O^{n}_{m}$ represent crystal field operators as discussed by 
Hutchings~\cite{Hutchings}, while $\alpha_{J}$,
$\beta_{J}$, and $\gamma_{J}$ are the Steven's coefficients~\cite{Stevens,Hutchings}. 
Trivalent Tb$^{3+}$ is an $^{7}\!F_{6}$ ion and thus the fixed $J$ manifold 
is $J=6$ ($L=3, S=3, J=L+S=6$)
for  the operator equivalent  point charge calculation.
The precise relationship between the point charge parameter set 
$\{\tilde B^k_q\}$ and the more general $\{B^{k}_{q}\}$ is discussed by Kassman~\cite{Kassman}.

Although a  point charge estimation of the crystal field parameters is in 
most cases unreliable  in  predicting the actual crystal field 
level spacing of rare earth ions, we find that varying the point charge 
parameter set  $\{\tilde B^k_q\}$ indicates that the ground state can be 
confirmed to be a 
competition between  two singlets close in energy and two magnetic doublets, 
which are well separated from the other crystal field states.
These levels  are indeed the 
remnants of the $\Gamma_{2}, \Gamma_{3}$ and $\Gamma^{(2)}_{5}$ states of the 
cubic environment eluded to earlier. 

In general, we find  that in a large region of the 
crystal field parameter space,  the two doublets form the lowest energy 
levels, although their precise ordering  may change.  Although the 
weight of each angular momentum component of a crystal field eigenstate 
varies with the values of the crystal field  parameters, some other general 
features do emerge. In particular, one singlet contains only $|\pm 6\rangle$ 
and $|\pm3\rangle$ states while the other is a combination of $|\pm 6\rangle,|\pm3\rangle$ and
$|0\rangle$. On the other hand both sets of  doublets are {\it magnetic}
(e.g they have a nonzero quantum expectation value of
$J^z$), and also contain components of exclusively different
$J^{z}$ values (the $J^z$ operator does not connect the
ground state to the excited state). One doublet has large 
$|\pm 4\rangle$ and $|\pm 1\rangle$ components, while the other has 
large $|\pm 5\rangle$ and $|\pm 2\rangle$  components. Hence, the
two doublets have $M_J$ components that differ by $\pm 1$, and a 
neutron spin-flip induced transition
from one to the other is allowed, consistent with what is found in
the inelastic neutron scattering results presented above. 

We have confirmed 
these conclusions based on our point charge analysis by performing a more 
sophisticated first principles calculation that take into account both
electrostatic and covalency effects as well as the intra atomic 
and configurational interactions. This approach, described in Appendix B,
does not restrict the decomposition
of the electronic energy levels into a {\it fixed} $\vert J,M_J \rangle$ manifold
as is usually done using the Steven's operator equivalents, as discussed above.
In the results
presented in Appendix B, we find that 
the two lowest energy doublets have a leading $M_J = \pm 4$ and $M_J \pm 5$
components, respectively (close to 90\% of the weight).
Table 2 in Appendix B lists three very similar energy level structures
given for slightly different constraints on the crystal field parameters.
The theoretically determined  energy levels
(mean values: 0, 13, 60 and 83 cm$^{-1}$, that is 0, 19, 86, and 119 K, respectively,
be compared with the experimental levels determined by inelastic neutron
diffraction (0, 17, 115 and 168 K). The corresponding wavefunctions allow for
a good estimate of the magnetic susceptibility of the diluted
and concentrated compounds as shown by Table 3 and Figs. 2, 3, and 8.

From these results, a picture of the low temperature single-ion properties of 
Tb$^{3+}$ can be deduced. Considering  the structure of the eigenstates of the two 
lowest doublets, a calculation of their $g-$tensors indicates extremely strong 
Ising like anisotropy along the appropriate $(111)$ axes (the axis formed by 
joining the two centers of the tetrahedra that the  ion belongs to) for each 
Tb$^{3+}$ ion at low temperatures. In summary, based on both point charge and
ab inition calculations, we have strong evidence of a doublet$-$doublet scheme
for Tb$^{3+}$, and 
a substantial single-ion anisotropy in ${\rm Tb_{2}Ti_{2}O_{7}}$~\cite{tb2ti2o7} 
making the the Tb$^{3+}$ moment effectively Ising$-$like for $T\lesssim O(10^1)$ K.
The consequences of this result will be discussed in Section IV.

\subsection{Susceptibility}

The zero field susceptibility measurements  of
the powder Tb-diluted compound
(Tb$_{0.02}$Y$_{0.98}$)$_2$Ti$_2$O$_7$
shown in Fig. 3 enables us  to gain a
more concise understanding  of the nature of the  crystal field levels of
${\rm Tb}^{3+}$ in ${\rm Tb_{2}Ti_{2}O_{7}}$, and the
single ion properties of Tb$^{3+}$.
Indeed, the dilute concentration of \tb ions in
the system removes the effect of magnetic interactions and  in
principle leaves only crystal field contributions to  the magnetic
susceptibility. 
Having obtained strong evidence for a doublet$-$ doublet scheme for
the Tb$^{3+}$ ions we now proceed further and analyse the
d.c. magnetic susceptibility by constructing a 
phenomenological expression for the susceptibility based on a
van Vleck equation
for such a doublet$-$ doublet energy level structure~\cite{van Vleck}.
This allows us 
to extract the size of the magnetic moments of both
doublets from experiment, as well as check for consistency
with the conclusions based on our crystal field calculations
of Appendix B.

Due to the powder nature of our 
(Tb$_{0.02}$Y$_{0.98}$)$_2$Ti$_2$O$_7$ sample, the random
orientation of the grains can lead to both parallel and transverse 
contributions to the susceptibility at first and second order. 
For the doublet-doublet structure at low temperature eluded to
earlier, the van Vleck equation for the susceptibility has the  
general form
\begin{equation}
\label{fit}
\chi=\frac{g^{2}\mu_{B}^{2}N_{\rm Tb}}{3k_{\rm B}}\left(\frac{a/T + b +e^{-\Delta/T}
(c/T - d)}{1 + e^{-\Delta/T}}\right),
\end{equation}
where $g$ is the Land\'{e} factor, equal to 3/2 for \tb, $\mu_{B}$ is the Bohr
magneton and $N_{\rm Tb}=0.04N_{0}$, where $N_{0}$ is Avogadro's constant. The 
adjustable parameters $a,b,c,d$ are defined through second order perturbation 
theory  to be 
\[
\begin{array}{cc}
a=\displaystyle{\sum_{\alpha,n_{0},m_{0}}}|\langle n_{0}|J^{\alpha}|m_{0}\rangle|^{2},
 &
c=\displaystyle{\sum_{\alpha,n_{1},m_{1}}}|\langle n_{1}|J^{\alpha}|m_{1}\rangle|^{2},
 \\
b=\displaystyle{2\sum_{\alpha,n_{0},m_{i \neq 0}}}\frac{|\langle n_{0}|J^{\alpha}|m_{i} 
\rangle|^{2}}{\Delta_{0,i}}, &
d=\displaystyle{2\sum_{\alpha,n_{1},m_{i \neq 1}}}\frac{|\langle n_{1}|J^{\alpha}|m_{i} 
\rangle|^{2}}{\Delta_{1,i}}, 
\end{array}
\]
where $n_{0},m_{0}$ label states within the ground state doublet, $n_{1},m_{1}$ 
label states in the excited doublet while the index $i$ defines any state from 
the $i$th crystal field level. The $\Delta$'s represent crystal field 
energy level 
differences (in Kelvin) and $\alpha=x,y$ or $z$. The fitting parameters $a$ 
and $c$ are due to first order terms in perturbation theory while $b$ and $d$ are
from second order terms  and give rise to
temperature independent van Vleck paramagnetism contributions to $\chi$.

We have performed a least squares fit to the susceptibility data of 
(Tb$_{0.02}$Y$_{0.98}$)$_2$Ti$_2$O$_7$ up to a temperature $\sim 30$ K, 
which is approximately where thermal contributions from a crystal field level 
at $\sim$ 100 K become non-negligible.  
Because of the narrow energy spacing between the ground state and excited state
doublet, {\it all four} adjustable 
parameters $a,b,c,d$ are important in fitting the susceptibility data. 
Based on the crystal field calculations of the previous section, the specific 
heat analysis of the doublet-doublet gap in the concentrated Tb$_2$Ti$_2$O$_7$ sample,
and the evidence of an anisotropy gap of $\sim$ 17 K 
observed in inelastic neutron measurements on the same sample~\cite{tb2ti2o7},
we have carried out the  
fit of the low temperature behavior of the susceptibility using  a 
doublet-doublet gap $\Delta$  ranging from
$[12 \sim 24]$ K.  We find that the goodness of fit is quite flat in 
this range for $\Delta$. However the magnitude of the adjustable parameters  
do not deviate strongly and can be 
determined to a reasonable degree of accuracy over this interval. Using values
of the gap outside of this interval yields a noticeably poorer goodness of fit. 
In Fig. 8, we show the fit to susceptibility data			
for the dilute (Tb$_{0.02}$Y$_{0.98}$)$_2$Ti$_2$O$_7$ sample 
using an anisotropy gap of 17 K as well as the results for the ab-initio crystal
field calculations.
						
We can interpret the fitted results for the susceptibility data by making use
of our crystal field results and Eq. (\ref{fit}).
Due to  the strong Ising like nature of the $g-$tensors of the two theoretically
calculated doublets, transverse terms such 
as $\langle n_i| J^\pm |m_i\rangle$ between two states {\it within} a doublet are  
negligible compared to $\langle n_i| J^z |n_i\rangle$. Thus, we expect that 
$a \simeq|\langle n_{0}|J^{z}|n_{0}\rangle|^{2}$  and 
$c\simeq|\langle n_{1}|J^{z}|n_{1}\rangle|^{2}$ and consequently, 
both the $a$ and $c$ terms represent permanent moment contributions to the 
susceptibility.  For the same reason, the $g-$tensor characterizing the
ground state is extremely anisotropic with essentially only a $g_\parallel$
component along the local $\langle 111\rangle$ direction 
with very little $g_\bot$ component.
As a result, and this is the most important point of the paper:
\begin{center}
{\it
At a temperature $T<O(10^1)$ {\rm K}, Tb$^{3+}$ ions can be considered to 
a very good approximation
as (classical) Ising magnetic moments confined to point parallel or antiparallel to
their local $(1 1 1)$  directions.
}
\end{center}
Over the range mentioned above for the doublet-doublet gap $\Delta$,
we have found that the magnitude of the moment in  the doublet 
ground state to be $5.10 \pm 0.3 \mu_{B}$. Overall, we find that our best fit 
for the susceptibility data gives $a=11.6 \pm .1, b= 1.53 \pm .04 \; 
{\rm K}^{-1},c=15.7 \pm 4.0$  and $d=.71 \pm .05\; {\rm K}^{-1}$. The value of 
$c$ gives a magnitude for the moment of the first excited doublet of 
$5.9 \pm .8 \mu_{B}$. 

Both fitted moments are compatible with the
eigenstate structures of the doublets  determined from  the crystal field
calculations presented in Appendix B.
Additionally, the values of the paramagnetic terms $b$ and $d$
in the susceptibility are also
consistent with our crystal field results. In particular, from the
$J^{z}$ components of our calculated low energy doublet and singlet eigenstates,
there will be predominant contributions to $b$ and $d$ coming from  transverse angular
momentum matrix elements connecting
the two doublets, as well as additional contributions involving transverse
matrix elements between the doublets and the higher energy
singlet states at $\gtrsim 100$ K.
The doublet-doublet coupling will give equal (in magnitude) contributions to
$b$ and $d$ while coupling to the singlets will give further  positive
contributions to $b$ while reducing the value of $d$. This can be simply
understood in terms of the signs of the denominators for each of these
virtual excitation processes in our definitions of $b$ and $d$.

In summary, we are able to successfully fit the
(Tb$_{0.02}$Y$_{0.98}$)$_2$Ti$_2$O$_7$
susceptibility measurements at low temperature 
using a doublet-doublet picture consistent with
our crystal field calculations.
We find reasonable agreement in terms of calculated moments and paramagnetic
contributions between that expected from theory and our 
fitted values.  In particular, we find a permanent moment in the ground state of 
approximately $5.1\mu_{\rm B}$. This moment is {\it intrinsic} to the
Tb$^{3+}$ ion and is {\it not} driven by exchange and/or dipolar
interactions as occurs in LiTbF$_4$~\cite{litbf4}.
 This value is also compatible to
what is estimated from the limiting low-temperature muon spin relaxation rate $1/T_1$
found in Ref.~\cite{tb2ti2o7}, assuming a dipole coupling between a positive muon
$\mu^+$ bounded to
an oxygen at $\sim 2.5\AA$ away from a $\sim 5\mu_{\rm B}$ Tb$^{3+}$ moment (see Appendix C).
\begin{figure}
\begin{center}
  {
  \begin{turn}{0}%
    {\epsfig{file=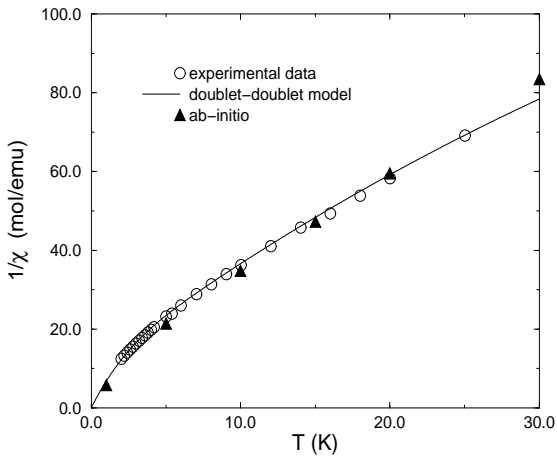,height=7.0cm,width=8.0cm} }
   \end{turn}
   }
\caption{Inverse susceptibility of (Tb$_{0.02}$Y$_{0.98}$)$_2$Ti$_2$O$_7$.
Circles indicate measured values while the solid line is
a fit using Eq. (\ref{fit}) using a doublet-doublet gap
of $\Delta=18$ K. The solid triangles are the values obtained from
the theoretical calculations of Appendix B.}
\end{center}
\end{figure}


\section{Discussion}						

Combining together our results from d.c. 
susceptibility measurements, specific heat data, inelastic neutron
scattering data, and crystal field calculations, the following picture emerges:

\begin{itemize}						
	\item The Tb$^{3+}$ ion in Tb$_2$Ti$_2$O$_7$ carries a 
		permanent magnetic moment of approximately 
		$5  \mu_{\rm B}$ . That moment is {\it intrinsic} to the
		Tb$^{3+}$ ion and is not driven by  magnetic correlation 
		from exchange and/or dipolar interactions.	
	\item	The ground state is well described as an Ising		
		doublet with extremely anisotropic $g-$tensor. In other words, the
		moments in the single-ion ground state are predominantly confined to
		point along the local $(111)$  directions.
	\item The first excited state is at an energy of approximately 18 $\pm 1$ K
		above the ground state, which is also characterized by a very anisotropic Ising-like $g-$tensor.
	\item Consequently, in the absence of interactions, 
		the Tb$^{3+}$ ions should be very
  		well modeled below $T \lesssim O(10^1)$ K as {\it effective}
		classical Ising spins confined to point along the local $(111)$ directions.  
\end{itemize}

From d.c. susceptibility measurements at high temperature, we know that
the magnetic interactions are predominantly antiferromagnetic,
with $\theta_{\rm CW}^{\rm exchange} \approx -13$ K. Although the
interactions are not
small compared to the first excitation energy gap of 18 K,  let us
momentarily ignore the exchange coupling between the ground state
and the excited doublet, and consider only the Ising-like ground state
doublet. We then have a classical model with effective Ising spins
pointing along their local $( 111)$ directions.
In other words, spins that can only point either inward or outward on a
tetrahedron.

Because of the open structure of the pyrochlore lattice, we expect that
the nearest-neighbor exchange interactions predominate. If we make the
further approximation that only nearest-neighbors contribute to the
antiferromagnetic interactions, we arrive at a scenario where
Tb$_2$Ti$_2$O$_7$ is effectively described by classical Ising spins
pointing along the $(111)$ directions and coupled via nearest-neighbor
antiferromagnetic exchange. As discussed in
Refs.~\cite{bramwell-ice,bramwell-obdo,moessner-cef},
such a model 
is {\it not} frustrated.
Indeed, if we pick one spin, and chose either an ``in'' or ``out''
orientation for it, then the three other spins on the same tetrahedron
must also take the same ``in'' or ``out'' configuration to minimize the
exchange interactions. Since the pyrochlore
lattice can be described as an FCC lattice 
with a tetrahedron
as the basis cell, either an ``all in'' or ``all out'' configuration repeats
identically on all basis units; one refers to such a magnetically long-range
ordered structure 
as a $Q=0$
ground state~\cite{reimers-mft,hertog-ice}.

The observation that Tb$_2$Ti$_2$O$_7$ remains paramagnetic down to at least 70 mK, while
sustaining important short range magnetic correlations, is therefore very
puzzling.  We believe that the most plausible explanation 
is that the interactions in
Tb$_2$Ti$_2$O$_7$ involve further than nearest-neighbor interactions,
$J_{nn}$. The presence of these interactions, as well as the long-range
dipolar interactions, possibly ``reintroduce''
large frustration in the spin Hamiltonian describing Tb$_2$Ti$_2$O$_7$,
 and conspire to destroy the 
``would-be'' long-range N\'eel $Q=0$ ground state, and give rise to a
collective paramagnet$-$spin liquid, 
ground state.


\section{Conclusion}

In conclusion, we have presented results from d.c. susceptibility, specific heat, inelastic neutron scattering, and
crystal field calculations for the pyrochlore lattice antiferromagnet Tb$_2$Ti$_2$O$_7$.
We have obtained strong evidence that the Tb$^{3+}$ magnetic 
moment on the 16d site at $T<2$ K is 
{\it intrinsic} and is not induced by magnetic   
(exchange and/or dipolar) interactions or correlation effects such as found in LiTbF$_4$~\cite{litbf4},  and is
of the order of $5 \mu_{\rm B}$ . All evidence points towards a very strong Ising like 
anisotropy for the doublet ground state which forces the resulting classical Tb$^{3+}$ Ising
moments to point either parallel or antiparallel to their local $\langle 111 \rangle$ direction.
 For antiferromagnetic
exchange interactions, such strong anisotropy largely removes all local ground state spin degeneracy,
and should naively force the system to possess an Ising-like long range order ground state with all spins
in or out on a tetrahedron basis cell. The reason for the failure of Tb$_2$Ti$_2$O$_7$ to order at a temperature
$\sim [10^0-10^1]$ K
set by the exchange part of the Curie-Weiss temperature remains unresolved at this time.
The results presented here point towards the need to consider the role that exchange interactions beyond
nearest-neighbor and dipolar interactions play in Tb$_2$Ti$_2$O$_7$.


\section{Acknowledgements}

It is a pleasure to acknowledge contributions from B. Canals, S. Dunsiger, 
R. Kiefl and Z. Tun to these studies.  We thank S. Bramwell,  B. Ellman, P. Holdsworth, 
and D. Schmitt for very useful and stimulating discussions. 
This research was funded by NSERC of Canada via operating grants and also			
under the Collaborative Research Grant, 
{\it Geometrically Frustrated Magnetic Materials}. M.G. acknowledges
the Research Corporation for a Research Innovation Award and a Cottrell
Scholar Award, and the Province of Ontario for a Premier Research
Excellence Award.

\begin{center}
{\bf APPENDIX A:\\
Dipole Contributions to $\theta_{\rm CW}$}
\end{center}

The contribution from dipolar interactions to $\theta_{\rm CW}$ may be determined 
from a high temperature series expansion of the long range dipole-dipole 
Hamiltonian to first order in $\beta$. That is to say, given  the additive 
property of crystal field effects, exchange and dipolar interactions to 
$\theta_{\rm CW}$, {\it i.e.} $\theta_{\rm CW}=\theta^{\rm cf}_{\rm CW} + \theta^{\rm ex}_{\rm CW} + 
\theta^{\rm dip}_{\rm CW}$, we may determine $\theta^{\rm dip}_{\rm CW}$ from an expansion of 
\begin{eqnarray*}
\chi=\frac{g^{2}\mu_{B}^{2}}{kT}\langle \sum_{ij}J^{z}_{i}J^{z}_{j} \rangle _{H^{\rm dip}}
\end{eqnarray*}
where $z$ defines some global direction and we assume $T\gg \theta^{\rm cf}_{CW}$,
{\it i.e.} isotropic spins. The angular brackets reflect an 
expectation value with respect to the dipole-dipole Hamiltonian $H^{\rm dip}$ 
defined by
\begin{eqnarray*}
H^{\rm dip}=\frac{1}{2}g^{2}\mu_{B}^{2}
\sum_{ij}\left(\frac{{\bf J}_{i}.{\bf J}_{j}}{|{\bf r}_{ij}|^{3}} - 
\frac{3({\bf J}_{i}\cdot {\bf r}_{ij})({\bf J}_{j}\cdot {\bf r}_{ij})}{|{\bf r}_{ij}|^{5}}
\right).
\end{eqnarray*}
To first order in $\beta$ this yields 
\begin{eqnarray*}
\chi=\frac{g^{2}\mu_{B}^{2}NJ(J+1)}{3kT}\left( 1 -
\frac{g^{2}\mu_{B}^{2}J(J+1)}{3NkT}\Lambda \right) ,
\end{eqnarray*}
and consequently that
\begin{eqnarray*}
\theta^{\rm dip}_{\rm CW}= - \frac{g^{2}\mu_{B}^{2}J(J+1)}{3Nk}\Lambda \;\;\; , \hspace{3cm} {\rm (A.1)}
\end{eqnarray*}
where
\begin{eqnarray*}
\Lambda=\sum_{ij}\frac{1}{|{\bf r}_{ij}|^{3}} - 
\frac{3|r^{z}_{ij}|^{2}}{|{\bf r}_{ij}|^{5}} 
\;\;\; .
\hspace{3cm} {\rm (A.2)}
\end{eqnarray*}

We see from the  above analysis  that the 
evaluation  of $\theta^{\rm dip}_{\rm CW}$ involves the  summation  of a  
conditionally convergent series (a lattice sum),
and thus special care must be taken in its 
evaluation. In general, it can be treated in a  controlled manner by the use 
of a rapid convergence factor via the Ewald method \cite{Born,Leeuw}.
Within  the Ewald approach, the  
sum is split into two rapidly converging sums, one over the real space lattice
and one in reciprocal space. Additionally, a surface (shape dependent) term 
also arises (which is interpreted as  a demagnetization factor\cite{Leeuw,Deem}).

Indeed, if we approximate the powder grains of the 
(Tb$_{0.02}$Y$_{0.98}$)$_2$Ti$_2$O$_7$
 as spherical, then the sum Eq. (A.2)
is identically zero. This can be understood by imagining Eq. 
(A.2)  as 
a sum over parallel (along the $z$ axis) dipoles moments of magnitude unity. 
For a spherical 
object, it can be  shown that such a sum must be identically zero for a 
system with  cubic symmetry\cite{Lorentz,Aharony}. Accordingly there would be no 
dipolar contribution 
to the measured value of $\theta_{\rm CW}$\cite{ewald-note}. On the other hand, we do
not expect that the geometry of the powder grains is in fact spherical, 
(additionally 
there is also the possibility of  effects from inter-granular interactions). 
Therefore to gain an estimate on the {\it upper bound} of the dipolar contribution,
we carry out the lattice sum Eq. (A.2)  for an infinitely long cylinder (needle shape) 
along the $z$ direction where surface effects are zero. This allows us to gain
an approximate upper bound on the dipolar contribution as there are 
effectively no demagnetization effects. This calculation can
be carried out rather simply by noting that for a spherical sample, Eq. (A.2),
can be written using
the Ewald method as~\cite{Leeuw}:
\begin{eqnarray*}
\Lambda_{\rm sphere}&=& \Lambda^{bulk} + \Lambda^{surface}_{\rm sphere} \;\;\; ,
\hspace{3cm} {\rm (A.3)}
\\
\noindent {\rm where}& & \\
\Lambda^{bulk}	&=& \frac{M}{L^{3}}\sum_{i\neq j}\left[
\sum_{\bf n}\frac{\alpha H(\alpha|\Rij|) +
(2\alpha/\sqrt\pi) e^{-\alpha^{2}|\Rij|^{2}}}{|\Rij|^{2}} \right. 
\nonumber \\
&-&\sum_{\bf n}\frac{3\alpha|{\bf R}^{z}_{ij}(n^{z})|^{2}
H(\alpha|\Rij|)e^{-\alpha^{2}|\Rij|^{2}}}{|\Rij|^{4}} \nonumber \\
&-&\sum_{\bf n} \frac{(2\alpha/\sqrt\pi)|{\bf R}_{ij}^{z}(n^{z})|^{2}
(3 + 2\alpha^{2}|\Rij|^{2})e^{-\alpha^{2}|\Rij|^{2}}}{|\Rij|^{4}} \nonumber \\
&+&\left.\sum_{{\bf n}\neq 0}
4\pi (n^{z}/|{\bf n}|)^{2}e^{-\pi^{2}|{\bf n}|^{2}/\alpha^{2}}e^{2\pi i 
{\bf n}.{\bf r}_{ij}/L} \right] \nonumber \;\;\;	,		
\;
{\rm (A.4)}
\\
\Lambda^{surface}_{\rm sphere} &=&\frac{M}{L^{3}}\sum_{i\neq j} \frac{4\pi}{3}  \;\;\; . \hspace{3cm} {\rm (A.5)}
\end{eqnarray*} 
The ${\bf R}_{ij}({\bf n})={\bf r}_{ij}/L + {\bf n}$, $i,j$ label the 
dipole (${\rm Tb}^{3+}$) sites within the cubic unit cell, $L$ is the length 
of 
the conventional cubic cell for 
Tb$_2$Ti$_2$O$_7$  and  $M$ is the number of cells in 
the sample. The function
$H(y)=(2/y\sqrt\pi)\int^{\infty}_{y}e^{-x^{2}}dx$ while ${\bf n}=(k,l,m)$ such 
that $k,l,m$ are integers and $\alpha$ is suitably chosen to ensure rapid
convergence of the summations over ${\bf n}$. 
Based on the previously mentioned  result  that 
$\Lambda_{spher}$ should vanish for a spherical sample, this implies 
that the first four terms 
in the above expression, $\Lambda^{bulk}$, must sum to the opposite 
value of the last term ($\Lambda^{surface}_{\rm sphere}$, the  surface term), and
indeed we have verified that this is the case numerically. For an infinitely 
cylindrical sample, the surface term, $\Lambda_{\rm cylinder}^{surface}=0$,  and thus one finds 
\begin{eqnarray*}
\Lambda_{\rm cylinder}& = &    \Lambda^{bulk} + \Lambda_{\rm cylinder}^{surface} \nonumber \\
		      & = &    \Lambda^{bulk}	\nonumber	\\
		&  = &   -\frac{M}{L^{3}}\sum_{i\neq j} \frac{4\pi}{3} \nonumber \\ 
		&\sim&  -1005 \frac{M}{L^{3}} \;\;\; .  \hspace{3cm}  {\rm (A.6)}
\end{eqnarray*}
On the other hand, for a slab geometry, we have
\begin{eqnarray*}
\Lambda_{\rm slab}& = & \Lambda^{bulk} +  \Lambda_{\rm slab}^{surface} \;\;\; .  \hspace{3cm} {\rm (A.7)}
\end{eqnarray*}
where
\begin{eqnarray*}
       \Lambda_{\rm slab}^{surface} & = & \frac{M}{L^{3}}\sum_{i\neq j} {4\pi}	
\;\;\; ,
\end{eqnarray*}
thus arriving at
\begin{eqnarray*}
\Lambda_{\rm slab}& = &  \frac{M}{L^{3}}\sum_{i\neq j} \frac{8\pi}{3}	
\;\;\; . \hspace{3cm} {\rm (A.8)}
\end{eqnarray*}
Combining Eqs.(A.1), (A.6) and (A.8), and using a cubic cell length of 
$L\sim {\rm 10.104 \AA}$, \cite{tb2ti2o7} we arrive at upper and lower bounds
for
dipolar contributions to the Curie-Weiss temperature, namely
$-2.4 \lesssim  \theta_{\rm CW}^{\rm dip} \lesssim +1.2$ K. 


\newpage

\vspace{2cm}
\begin{center}
{\bf APPENDIX B:\\
 Crystal Field Calculations}
\end{center}

Our aim in this Appendix  is to pursue in more details and using a more sophisticated
(ab-initio) approach the calculations of the magnetic susceptibilites of
{\tbtio} and {\ytio}.
 To do so, we need
to determine the electronic structure of the ground {\FS} level of
the {\tbt} ($4f^8$ configuration) in these two compounds, deduce the 
wavefunctions and from there infer the value of the magnetic moment of the ground 
state level.
The determination of the $4f^8$ electronic configuration is obtained by 
diagonalizing  the  following Hamiltonian for a generic $f^n$ 
configuration without making the
fixed $J$ manifold approximation used in Section  III.A:
\begin{eqnarray*}
H(f^n) & = & \sum_k F^k(ff)f^k  + \zeta(f) A_{so} + \alpha L(L+1) + 	\\
 &  &\beta C({\rm G_2}) +
 \gamma C({\rm R_7}) +  \sum_{kq} B_q^k\;C_q^k  \;\;\; . \hspace{1.5cm} {\rm (B.1)}
\end{eqnarray*}
Let us explain first what are the different terms in  $H(f^n)$.

\begin{itemize}
\item The $F^k$'s ($k=2,4,6$) are the electrostatic integrals (Slater's
parameters) which splits the $4f^n$ configurations into terms $^{2S+1}L$
where $S$ is the total spin and $L$ is the total orbital angular
momentum.
The $f^k$'s are the associated two-electron operators~\cite{judd-98}.
\item  $\zeta(f)$ is the spin-orbit interaction integral  which splits the
terms into  $^{2S+1}L_J$ levels. $A_{so}$ is the
associated one-electron spin-orbit operator~\cite{judd-98}.
\item  $\alpha$, $\beta$ and $\gamma$  are parameters associated with effective
two-body correction terms for inter-configuration interaction~\cite{rajnak}.
$C$(G$_2$) and $C$(R$_7$) are the Casimir's operators for
groups G$_2$ and R$_7$.
When $2<n<12$, there can be several terms $^{2S+1}L$ with the same
$S$ and $L$ values in the f$^n$ configuration. For instance there are three
$^5$G terms in 4f$^8$ while there exists only one term
$^7$F. The states may differ by the way they are built from the parent
configuration f$^{n-1}$. An additional classification of the states is
therefore necessary. It is done according to the irreducible representations
of the the groups G$_2$ and  R$_7$ and bestows additional quantum numbers to
the states.
\item the $B_q^k$ are the coefficients of the one-electron crystal field
interaction which acts between $\vert ^{2S+1}LJM_J\rangle$ sublevels. They can be
theoretically predicted or
extracted from fits of the energy  levels  (spectral lines)
from experiments. In the
point charge electrostatic model, their expression is :
\begin{eqnarray*}
B_q^k = (4\pi/2k+1)^{1/2}  \langle r^k\rangle \Sigma_j (Q_j/R_j^{k+1})
 Y_k^{q*} ( \theta _j \phi _j )	
\end{eqnarray*}
Where $\langle r^k \rangle$ is a $4f$ electron radial integral, $ Q_j $ is the point
charge of ligand $j$, and  $R_j$,  $\theta_j$, and  $\phi_j$, are the
polar coordinates of ligand $j$. The derivation of
the $B_q^k$ for the covalent interactions is much more involved\cite{garcia-85}.
The $C_q^k $ =  (4$\pi$/2k+1)$^{1/2}[[ { \bf Y}_k^{q} ]] $, are the
tensorial  one-electron  crystal field operators.
\end{itemize}

 The evaluation of the matrix elements of {\it i)} 
the electrostatic interaction,
{\it ii)} the spin-orbit interaction, {\it iii)}
the free-ion configuration interaction,
and {\it iv)} the crystal field interaction, between the
states of the basis set chosen for the  $f^n$ configurations 
is necessary in order to determine the eigenvalues and eigenvectors 
of the latter. The matrix elements are calculated
by the means of tensorial algebra~\cite{judd-98}.
Besides, if coupled $\vert ^{2S+1}LJM_J\rangle $ states are chosen as the basis set,
the  one- or two-electron operators which are involved in the Hamiltonian
cannot act directly on them. The calculation requires intermediary 
mathematical quantities known as reduced matrix
elements which are tabulated for standard  $f^n$ configurations
(~\cite{nielson}). Once evaluated, the complete matrix elements are
multiplied by the associated parameters before diagonalization. The 
parameters are then determined by trial and error by successive 
diagonalizations and comparison of the eigenvalues with experimental 
energy levels.
In the present case our main concern is the structure of the Tb$^{3+}$ 
ground level, so that the only specific material dependent parameters
which have to be determined before diagonalizing $H(f^n)$ are
the $B_q^k$ crystal-field interaction parameters.
As already mentioned , the most convenient way to deduce the electronic 
structure of a rare earth ion in a solid compound, 
and hence determine the crystal-field parameters, is usually via
analysis of its electronic spectrum by fitting the $B_q^k$ 
to match the frequency of the observed transitions. The
strongest lines are due to electronic transitions partly
allowed by the mixing of the ground configuration with opposite parity
configurations (Judd-Ofelt mechanism)~\cite{judd,ofelt}.
However, in the case of pyrochlores, the
site symmetry at the rare earth site is centrosymmetrical. The odd parity
crystal field parameters vanish, the mixing of opposite parity
configurations is impossible, hence no electric dipole transitions are detectable 
in the spectrum. Indeed, only the  weak  {\FU}$\to${\DO} and
{\FO}$\to${\DU}
magnetic dipole transitions were observed previously in the absorption spectra
of the pyrochlore compounds {\eutio} and {\eusno}~\cite{caro}.
Hence a complete set of ``phenomenological"
crystal field parameters (CFP) cannot be determined from optical absorption or emission spectra.
However, recent inelastic neutron scattering experiments have been able to give information
on the lowest electronic levels of Ho$_2$Ti$_2$O$_7$~\cite{shastry}. From these
neutron results, some CFP's can be deduced. CFP's can also be calculated 
ab-initio from the compound structure and the atomic data of the constituents.
Therefore, in what follows, two approaches are  used  to determine the
CFP's of 
Tb$_2$Ti$_2$O$_7$ and (Tb$_{0.02}$Y$_{0.98}$)$_2$Ti$_2$O$_7$:

\newcounter{bean2}
\begin{list}%
{\bf {\Roman{bean2}A)}}{\usecounter{bean2}\setlength{\rightmargin}{\leftmargin}}
\item
A full ab-initio calculation of the CFP's utilizing the structural data of the compounds.
\end{list}

\newcounter{bean1}
\begin{list}%
{\bf {\Roman{bean1}B)}}{\usecounter{bean1}\setlength{\rightmargin}{\leftmargin}}
\item 
The fit of  the CFP's from the Ho$_2$Ti$_2$O$_7$~\cite{shastry} and
Eu$_2$Ti$_2$O$_7$~\cite{caro} data and a
transposition to the Tb$_2$Ti$_2$O$_7$ and (Tb$_{0.02}$Y$_{0.98}$)$_2$Ti$_2$O$_7$
compounds.
\end{list}

\vspace{1mm}

The next two steps for the calculations of the magnetic susceptibility are:

\begin{list}%
{\bf{\Roman{bean1})}}{\usecounter{bean1}\setlength{\rightmargin}{\leftmargin}}
\addtocounter{bean1}{1}
\item The calculation of the 4$f^8$ electronic configuration utilizing plausible
free ion parameters, and the fitted or calculated ab-initio values of the CFP's.
\item The calculation of the magnetic susceptibility of
Tb$_2$Ti$_2$O$_7$ and (Tb$_{0.02}$Y$_{0.98}$)$_2$Ti$_2$O$_7$
utilizing the wavefunctions derived from the previous step II.
\end{list}

\vspace{2mm}

These steps are detailed in the following :

\vspace{2mm}

\noindent {\bf IA)} Ab-initio calculation of the CFPs
\vspace{1mm}

An ab-initio determination of the CFP's is obtained by adding an electrostatic and a
covalent contribution along lines similar to the ones developed in 
Refs.~\cite{garcia-85,garcia-92} for oxygen ligands.
The crystal structure, the ionic charges and the
ionization energies of the ligands are used. In Ref.~\cite{garcia-95},
experimental and predicted values of the parameters calculated by the
``covalo-electrostatic" model were compared for ten compounds. The mean 
deviation between experimental and calculated values:  
\begin{eqnarray*}
\Delta B^{k}/B^{k} & = & \left[ \sum_{-k \leq {q} \leq k} (B^{k}_{qe}-B^{k}_{qc})^2/
 \sum_{-k \leq {q} \leq k} (B^{k}_{qe})^2 \right] ^{1/2}  
\\
		& = & [1+1/(S^k)^2-2\cos (R^k)/S^k]^{1/2}
\end{eqnarray*}
where $S^k$ and $R^k$ are the scale and
reliability factors listed in Table 7 of Reference~\cite{garcia-95}.
$\Delta B^{k}/B^{k}$
is found to be equal to 52, 30 and 23\% for $k=2$, $k=4$ and $k=6$, 
respectively. Such is the uncertainty which can be expected from
a ``blind eyed" prediction of the CFP's of \tbtio.

As mentioned earlier, the space group of rare earth titanates with
the pyrochlore structure is Fd$\bar 3$m. The eight oxygen
first neighbours form a distorted cubic polyhedron. Two oxygens occupy ideal
positions on opposite summits of the cubic threefold axis. The three [[ sides ]]
of the cube originating from each of these two summits are equally
elongated. The $a$ cubic lattice parameter is equal to 10.15 and 10.09
$\AA$, and $x$ the positional parameter for the six displaced oxygens 
to 0.3 and 0.2968 for the dense and dilute compound respectively~\cite{knop}.

The site symmetry at the rare earth site is reduced from O$_h$ (cubic) to
D$_{3d}$. The remaining threefold order symmetry axis imposes for 
the crystal field parameters the condition $q=0$, modulo 3, so that the
non-zero even $k$ CFP's are 
{\bzd}, {\bzq}, {\btq}, {\bzs}, {\bts} and {\bss}. The 
predicted CFP values are reported in Table 1. The distances
between the Tb$^{3+}$ ion and the oxygens on the threefold axis is short (2.20 
and 2.18 $\AA$ for the dense and the dilute compound, respectively) while the
distances to the six peripheral oxygens are much larger (2.52 and 2.49 $\AA$,
respectively). This explains why the ``axial" {\bzk} parameters are much
larger than the ``azimuthal" \bkq's .

\vspace{2mm}

\noindent {\bf IB)} Experimental determination of the CFPs.

\vspace{1mm}

The CFPs can be determined by fitting the energy levels of
the $f^n$ configuration either using spectroscopic or inelastic neutron scattering
data from other pyrochlore compounds. Here, we can 
use the {\bzd}  from spectroscopic data on
Eu$_2$Ti$_2$O$_7$~\cite{caro} and the other $B_q^k$ from neutron scattering data on
Ho$_2$Ti$_2$O$_7$~\cite{shastry}.

\vspace{1mm}

\begin{itemize}

\item Determination of $B_0^2$ from spectroscopic data on Eu$_2$Ti$_2$O$_7$ \newline

The {\FU}$\to${\DO} and {\FO}$\to${\DU} magnetic dipole transitions were observed
previously in the electronic absorption spectrum of {\eutio}~\cite{caro}.
In the latter,
the {\FU} and {\DU} splittings amount to 
291 K and 51.9 K, respectively. A fit
in {\qfs}({\Eu}) yields  {\bzd} = 684 K. The transposition to  {\tbt} is made
assuming the crystal field parameter is scaled by the ratio of radial
integrals:
$B_0^2$(Tb$^{3+}$) = $B_0^2$(Eu$^{3+}$)
$\times$$\langle r^2 \rangle$(Tb$^{3+}$)/$\langle r^2 \rangle$(Eu$^{3+}$)
= $B_0^2$(Eu$^{3+}$) $\times$ 0.91 = 622 K.
This ``experimental'' value, listed in Table 1, and
 referred  to as (a), has the same sign, but it is about half
of the predicted value. As pointed out hereabove, the
uncertainty on the calculated $k=2$ parameters is large.

\item CFP determination in Ho$_2$Ti$_2$O$_7$.  \newline

Siddharthan et al.~\cite{shastry} recently reported results from ineleastic neutron
scattering at low temperature in Ho$_2$Ti$_2$O$_7$.
They determined the six lowest E irreducible representations of the $^5$I$_6$ 
ground state level. Utilizing their experimental values, we fitted the
CFP's of Ho$^{3+}$ while maintaining the ratio between
CFP's with the same $k$ value close to the theoretical ratio. 
The program ATOME was used for the refinement~\cite{ATOME}. In this program, the basis
set is composed of Slater determinants which makes unnecessary the use of tables of
reduced matrix elements~\cite{ATOME}. The evaluation of the matrix elements
is straightforward, but the configuration cannot be truncated. Indeed,
each eigenvector being a linear combination of a large number of
Slater determinants, none of the latter can be omitted.
As a consequence, all the 1001 states of 4f$^{10}$ configuration are included
in the diagonalization matrix.  The basis is large (1001) but
still tractable.  The final mean
deviation between experimental and calculated levels was equal to 7.8 K.
The fitted CFP's are : \bzq = 3173 K, \btq = -1459 K, \bzs =
1343 K, \bts = 1292 Kand \bss = 609 K. As pointed out hereabove, the CFP's were then
scaled according to the ratio between the radial integrals of Ho$^{3+}$ and Tb$^{3+}$
to give the experimental CFP's for Tb$_2$Ti$_2$O$_7$.
Namely, $B_q^k$(Tb$^{3+}$) = $B_q^k$(Ho$^{3+}$)
$\times$$\langle r^k \rangle$(Tb$^{3+})$/$\langle r^k \rangle$(Ho$^{3+}$).
These ``experimental values'' are
are listed in Table 1, and  referred  to as (b).
The experimental $k=4$ and $k=6$ CFP's
are 1.7 and 1.9 times larger than the theoretical values  which is somewhat unusual.

\end{itemize}

\noindent {\bf III)} Calculation of $4f^8$ electronic configuration.

\vspace{1mm}

The calculation of the $4f^8$ electronic configuration is done by the
means of program $f^n$ ~\cite{faucher-code}
utilizing the Hamiltonian $H(f^n)$ in B.1
acting on coupled states $^{2S+1}$L$_J$. 
Contrary to program ATOME previously mentioned, program f$^n$ can work
on a truncated basis, which is necessary to resolve the \qfh configuration
of \tbt with a large number of states (3003 states).  
The interaction matrix is built on a 387x387 basis set comprising the
following  $^{2S+1}$L terms of the \tbt ( \qfh ) configuration: $^7$F,
$^5$D(1,2,3), $^5$F(1,2), $^5$G(1,2,3), $^3$P(1,2,3,4,5,6) and
$^1$S(1,2,3,4). The conjugate configuration of {\tbt} is that of {\Eu} with
$4l+2-8=6$ electrons. The $4f^6$ (n=6) configuration of Eu$^{3+}$
contains exactly the same
number of basis states (e.g. [4$l$+2]!/[[4$l$+2-n]!n!] = 3003 states) as {\tbt},
the same terms, and the same
levels.
The interactions involving an odd number of electrons have a reverse sign in
4f$^6$ and 4f$^8$.  For instance the  $^{2S+1}$L$_J$ levels appear in a reverse
order, and so do the crystal field sub-levels. In addition, the terms
determined by the electrostatic interaction (two electron interaction)
appear up in the same order for {\Eu} and {\tbt}.

For {\Eu} the above quoted basis had proved large enough to allow 
a simulation of the levels up to
$^5$D$_2$ (30219 K)  without drastic truncation effects~\cite{moune}.
The  $F^k$'s were assigned the Gd$^{3+}$
values given in Ref.~\cite{sovers}, that is 147289 K, [[ 102479 K ]], and
55868 K for $k=$ 2, 4 and 6, respectively. $\alpha$, $\beta$, $\gamma$
were ascribed the {\Nd} values fitted in Ref.~\cite{chertanov}, that is 
30.98 K, -1005.03 K, and 2510.48 K,  respectively.   
The spin-orbit coupling constant $\zeta(f)$ was set
equal to 2446.30 K which is a standard value for {\tbt}~\cite{carnall}.
{\bzd} was assigned the transposed value quoted hereabove, and the other CFP's
the values listed in Table 1 obtained after
rescaling the $B_q^k$ extracted from fits to the levels of 
Ho$^{3+}$. The diagonalization of the interaction matrix
gives the energy levels and the corresponding leading eigenvectors listed
in Table 2. 
The lowest levels are two doublets. Both states are rather Ising-like, with nearly
exclusive ($\sim 0.95$) $M_J=\pm 4$ and $M_J=\pm 5$ components. This is caused by the
very ``axial" crystal field parameters. In other words, the $M_J=\pm 4$ and
$M_J=\pm 5$ carry, respectively, about 90\% of the weight of the ground state and first excited
state doublet wavefunctions.

\newpage

\noindent {\bf III)} Calculation of the  magnetic susceptibility of
Tb$_2$Ti$_2$O$_7$ and {\ytio}.

\vspace{1mm}

The d.c. magnetic susceptibility is
by Van Vleck's formula~\cite{beaury,van Vleck}
using the eigenvectors determined in the previous step.
\begin{eqnarray*}
 \chi = \frac{ N \mu_{\rm B} ^2}{ \sum_i e^{-E_i^{(0)}/k_{\rm B}T}} 
\left[  \sum_i \frac { ( \epsilon_i^{(1)})^2}{k_{\rm B}T}-2 \epsilon_i^{(2)}
\right]  e^{-{E_i^{(0)}/k_{\rm B}T}}
\end{eqnarray*}
where $N$ is the number of moles of Tb$^{3+}$,
 $\mu_{\rm B}$ is the Bohr magneton, $k_{\rm B}$ is
the Boltzmann constant, $E_i^{(0)}$ is the energy of the $i^{th}$ level.
Besides,
\begin{eqnarray*}
 \epsilon_i^{(1)} = \langle \psi_i | \vec{L} + \vec{2S} | \psi_i \rangle  
\end{eqnarray*}
\begin{eqnarray*}
\epsilon_i^{(2)} = \sum_{j \neq i}
\frac{ ( \langle \psi_i | \vec{L} + \vec{2S} | \psi_j \rangle )^2}{E_i^{(0)}-E_j^{(0)}} 
\end{eqnarray*}
The results as a function of the temperature are given in Table 3. 
The values for  $\langle   \mu(T)\rangle$ are
very similar for the {\tbtio} and {\ytio} compounds. Therefore any
large experimentally stated difference between the two compounds at low
temperature cannot be accounted for by a difference in the individual
characteristics of {\tbt} in the dense {\tbtio} and the dilute {\ytio}.



\vspace{2.0cm}


\begin{center}
{\bf TABLE 1}\\
\end{center}
\noindent{
Empirical CFP's deduced by the covalo-electrostatic model for=
Tb$_2$Ti$_2$O$_7$
and  (Tb$_{0.02}$Y$_{0.98}$)$_2$Ti$_2$O$_7$, 
and ``experimental'' CFP's for
Tb$_2$Ti$_2$O$_7$. The latter are obtained using two approaches.
(a) refers to the value of $B_0^2$ obtained by transposing to Tb$^{3+}$ the
value of $B_0^2$ determined from spectroscopic data of Eu$_2$Ti$_2$O$_7$.
(b) refers to the values $B_q^{k\ge 4}$ obtained by transposing to Tb$^{3+}$ 
the
values of $B_q^{k\ge 4}$ determined from inelastic neutron data on  Eu$_2$Ti$_2$
O$_7$.
All values in K. (PCEM = point charge electrostatic model.)}

\begin{tabular}{lrrrrrrr}
         &           &  \bzd &  \bzq  &  \btq  &  \bzs &  \bts  &  \bss \\ 
         &    PCEM   &   471 &   708  &  -187 \\ 
\tbtio   &     Cov.  &   610 &  1599  &  -227  &   1261 &   314 &   482 \\ 
\\
         &     Total & \bf 1081 \rm & \bf 2307 \rm & \bf -414 \rm  
                     & \bf 1261 \rm & \bf  314 \rm & \bf  482 \rm \\
\\
         &    PCEM   &   407 &    731 &  -210 \\
\ytio    &    Cov.   &   609 &   1711 &  -288  &   1324 &   389 &   571 \\ 
\\ 
         &    Total  & \bf 1016 \rm & \bf 2442 \rm & \bf -498 \rm  
                     & \bf 1324 \rm & \bf  389 \rm & \bf  571 \rm \\
\\
``Experimental'' &      & \bf 622$^a$  \rm & \bf 3691$^b$ \rm & \bf -1698$^b$ \rm
                     & \bf 1731$^b$ \rm & \bf 1665$^b$ \rm & \bf   784$^b$ \rm \\  
\end{tabular}

\newpage

. \newpage

\vspace{3mm}
\begin{center}
{\bf TABLE 2}\\
\end{center}
\noindent{Lowest energy levels (in K), irreducible representations in 
(Tb$_{0.02}$Y$_{0.98}$)$_2$Ti$_2$O$_7$ and
Tb$_2$Ti$_2$O$_7$,
and leading compositions of the
corresponding eigenvectors. 
Lines labelled (a) and (b) are for
the {\it predicted} projections for the dilute
Tb$_{0.02}$Y$_{0.98}$)$_2$Ti$_2$O$_7$ and
dense Tb$_2$Ti$_2$O$_7$, respectively, using the CFP's listed in Table 1.
Lines labelled (x3) are for 
the predicted projections for the dense
Tb$_2$Ti$_2$O$_7$
using the ``experimental'' CFP's  listed in Table 1.
}

\normalsize

\begin{tabular} {llccccccccc}
 &   &  &  & &  & & &  & \\
(E)&  En &  $^7$F$_6$ , +/-4  & $^7$F$_6$ ,+/-1 & $^7$F$_6$,-/+5 & $^5$G$_6$(1), +/-4  & $^5$G$_6$(3) , +/-4 \\
(a)&  0. &  -0.97        &+/-0.06    & -/+0.06  &  +0.15         &   -0.14 \\
(b)&  0. &  -0.97        &+/-0.05    & -/+0.06  &  +0.15         &   -0.14 \\
(c)&  0. &  -0.95        &+/-0.13    & -/+0.13  &  +0.14         &   -0.13  \\
\\
(E)& En & $^7$F$_6$ , +/-5 & $^7$F$_6$ ,+/-2 & $^7$F$_6$,-/+4 & $^5$G$_6$(1), +/-5  & $^5$G$_6$(3) , +/-5 \\
(a)&18.7  &   -0.96      & +/-0.13      &  -/+0.07& +0.14         &-0.13   \\
(b)&15.9  &   -0.97      & +/-0.11      &  -/+0.06& +0.14         &-0.14   \\
(c)&21.6  &   -0.92      & +/-0.28      &  -/+0.14& +0.14         &-0.13   \\
\\
(A2)& En  & $^7$F$_6$,-3  & $^7$F$_6$,3 & $^7$F$_6$,-6 & $^7$F$_6$,6  & $^5$G$_6$(1),-3 &$^5$G$_6$(3),3 \\
(a)&86.5  & +0.67   &+0.67  &+0.16    & -0.16     &-0.10      &-0.10         \\
(b)&87.9  & +0.67   &+0.67  &+0.16    & -0.16     &-0.10      &-0.10         \\
(c)&85.0  & +0.65   &+0.65  &+0.22    & -0.22     &-0.10      &-0.10         \\
\\
(A1)& En  & $^7$F$_6$,-3  & $^7$F$_6$,3 & $^7$F$_6$,-6 & $^7$F$_6$,6  & $^7$F$_6$,0 \\
(a)&121.1  & +0.66   &-0.66  &+0.21   & +0.21  &+0.07              \\
(b)&118.2  & +0.66   &-0.66  &+0.20   & +0.20  &+0.06              \\
(c)&119.6  & +0.64   &-0.64  &+0.25   & +0.25  &+0.11         \\
\end{tabular}



\vspace{2.0cm}


\begin{center}
{\bf TABLE 3}\\
\end{center}
\vspace{3mm}
\noindent{
Ionic magnetic moment (in Bohr magnetons) and inverse molar magnetic
susceptibility as a function of temperature.
}

\vspace{1cm}

\begin{tabular}{lccc}
$T$ (K) & $\langle \mu \rangle$ ($\mu_{\rm B}$)  & $1/\chi$ (mole/emu) &  $1/\chi$ (mole/emu)   \\
        &    Tb$_{0.02}$Y$_{0.98}$Ti$_2$O$_7$   &  Tb$_{0.02}$Y$_{0.98}$Ti$_2$O$_7$ & Tb$_2$Ti$_2$O$_7$      \\
\\
 1 &  5.90 & 5.8  & 0.11  \\
 5 &  6.84 & 21.4 & 0.41  \\
 10 & 7.58 & 34.8 & 0.67  \\
 15 & 7.97 & 47.3 & 0.92  \\
 20 & 8.20 & 59.5 & 1.17  \\
 30 & 8.47 & 83.5 & 1.65  \\
 300 &9.22 & 706  & 14.09  \\
\end{tabular}

\newpage
.\newpage



\begin{thebibliography}{99}
\bibitem{spinel}Strictly speaking, ZnFe$_2$O$_4$ and ZnCr$_2$O$_4$ are B-site spinels
with no, or little inversion. However, the magnetic lattice of the B-site spinel is
identical to the lattice of corner-sharing tetrahedra in pyrochlores~\cite{villain}.
\bibitem{ewald-note}
Using the Ewald method, one finds that the  surface term cancels the
combined effect of both real and reciprocal space sums.
\end{thebibliography}
\end{document}